\documentclass{aa}  
\usepackage{graphicx}
\usepackage{upgreek}
\usepackage{amsmath}

\usepackage{txfonts}
\usepackage{color}
\usepackage{hyperref}
\hypersetup{%
  colorlinks=true,% hyperlinks will be black
  linkcolor=blue,% hyperlink borders will be red
  citecolor=blue
}

\def \bmath #1 {{\hbox{\boldmath{$#1$}\unboldmath}}}

\usepackage[dvipsnames]{xcolor}
\begin{document}

\bibliographystyle{aa}

\title{Gas-phase Elemental abundances in Molecular cloudS (GEMS)}
\subtitle{III. Unlocking the CS chemistry: the CS+O reaction}
\authorrunning{Bulut et al}
\titlerunning{CS+O reaction}
 
\author{
  Niyazi Bulut\inst{1}
  \and
  Octavio Roncero\inst{2}
  \and
  Alfredo Aguado\inst{3}
  \and
  Jean-Christophe Loison\inst{4}
\and 
  David Navarro-Almaida\inst{5}
   \and
   Valentine Wakelam\inst{6}  
   \and
  Asunci\'on Fuente\inst{5}
 \and
  Evelyne Roueff\inst{7}
  \and
  Romane Le Gal\inst{8}
 \and
  Paola Caselli\inst{9}
  \and
  Maryvonne Gerin\inst{7}
   \and
  Kevin M. Hickson\inst{6}
  \and
  Silvia Spezzano\inst{9} 
  \and
  P.~Rivi\'ere-Marichalar\inst{5}
  \and
  T. ~Alonso-Albi\inst{5}
  \and
  R.~Bachiller\inst{5}
   \and
  Izaskun Jim\'enez-Serra\inst{10}
  \and
  C.~Kramer\inst{11}
  \and
  Bel\'en Tercero\inst{5,12}
  \and
  Marina Rodriguez-Baras\inst{5}
     \and
   S.~Garc\'{\i}a-Burillo\inst{5}   
    \and 
  Javier R. Goicoechea\inst{2}
     \and
    S.~P. Trevi\~no-Morales\inst{13}
    \and
   G. Esplugues\inst{5}
   \and
  S.~Cazaux\inst{14} 
  \and
   B.~Commercon\inst{15}
 \and
 J. Laas\inst{9}
     \and
    J.~Kirk\inst{16}
    \and
    V.~Lattanzi\inst{9}
     \and
     R.~ Mart\'{\i}n-Dom\'enech\inst{8}
     \and
     G.~Mu\~noz-Caro\inst{10}
     \and
    J.~Pineda\inst{9}
    \and
    D.~Ward-Thompson\inst{16}             
     \and
    M.~Tafalla\inst{5}  
     \and
     N.~Marcelino\inst{2}                        
     \and
    J.~Malinen\inst{17,18}
    \and
   R.~Friesen\inst{19}
     \and
   B.~M.~Giuliano\inst{9}
 \and
 M.~Ag\'undez\inst{2}
 \and
   A.~Hacar\inst{20}
}

\institute{
%1
  University of Firat, Department of Physics, 23169
          Elazig, Turkey
       \and
%2       
          Instituto de F{\'\i}sica Fundamental (IFF-CSIC), C.S.I.C.,                                   
          Serrano 123, 28006 Madrid, Spain.
 %3         
          \and
          Departamento de Qu{\'\i}mica-F{\'\i}sica Aplicada, Unidad asociada IFF-UAM,
          Universidad Aut\'onoma de Madrid, 28049 Spain
          \and
%4          
          Institut des Sciences Mol\'eculaires ISM), CNRS, Univ. Bordeaux, 351 cours de la
            Lib\'eration, F-33400, Talence, France
          \and
%5
          Observatorio Astron\'omico Nacional (IGN), c/ Alfonso XII 3, 28014 Madrid, Spain.
          \and         
 %6                  
        Laboratoire d'astrophysique de Bordeaux, Univ. Bordeaux, CNRS, B18N,
        all\'ee Geoffroy Saint-Hilaire, 33615 Pessac, France
          \and
 %7         
        Sorbonne Universit\'e, Observatoire de Paris, Universit\'e PSL, CNRS, LERMA, F-92190,  Meudon, France
         \and
   %8       
        Center for Astrophysics | Harvard \& Smithsonian, 60 Garden St., Cambridge, MA 02138, USA
        \and
 %9
     Centre for Astrochemical Studies, Max-Planck-Institute for Extraterrestrial Physics,Giessenbachstrasse 1, 85748, Garching, Germany    
    \and
%10     
        Centro de Astrobiolog{\'\i}a (CSIC-INTA), Ctra. de Ajalvir, km 4, Torrej\'on de Ardoz,
        28850 Madrid, Spain     
        \and
 %11
     Instituto Radioastronom{\'\i}a Milim\'etrica (IRAM), Av. Divina Pastora 7, Nucleo Central, 18012, Granada, Spain 
     \and  
 %12      
          Observatorio de Yebes (IGN), Cerro de la Palera s/n, 19141 Yebes, Guadalajara, Spain.
       \and
%13
  Chalmers University of Technology, Department of Space, Earth and Environment, SE-412 93 Gothenburg, Sweden
  \and
%14
  Faculty of Aerospace Engineering, Delft University of Technology, Delft, The Netherlands ; University of Leiden, P.O. Box 9513, NL, 2300 RA, Leiden, The Netherlands 
\and
%15
  \'Ecole Normale Sup\'erieure de Lyon, CRAL, UMR CNRS 5574, Universit\'e Lyon I, 46 All\'ee d'Italie, 69364, Lyon Cedex 07, France 
\and
%16
 Jeremiah Horrocks Institute, University of Central Lancashire, Preston PR1 2HE, UK
\and
%17
 Department of Physics, University of Helsinki, PO Box 64, 00014 Helsinki, Finland
 \and
%18
 Institute of Physics I, University of Cologne, Cologne, Germany
 \and
 %19
 National Radio Astronomy Observatory, 520 Edgemont Rd., Charlottesville VA 22901, USA 
\and
%20
 Leiden Observatory, Leiden University, PO Box 9513, 2300-RA, Leiden, The Netherlands 
     }

\abstract{Carbon monosulphide (CS) is among the most abundant gas-phase S-bearing molecules in cold dark molecular clouds.
  It is easily observable with several transitions in the
  millimeter wavelength range, and has been
 widely used as a tracer of the gas density  in the interstellar medium  in our Galaxy and external galaxies. 
However, chemical models fail to account for the observed CS abundances when assuming the cosmic value for the elemental
abundance of sulfur. }
 {The CS+O $\rightarrow$ CO + S reaction has been proposed as a relevant CS destruction mechanism at low temperatures, and could explain the discrepancy between models and observations. Its reaction rate has been experimentally measured at temperatures of 150$-$400~K, but the extrapolation to lower temperatures is doubtful. Our goal  is to calculate the CS+O reaction rate at temperatures $<$150 K which are prevailing in  the interstellar medium.}
 {We performed {\it ab initio} calculations
   to obtain the three lowest potential energy surfaces (PES) of the CS +O system. These PESs are  used to study the reaction dynamics,
   using several methods (classical, quantum, and semiclassical) to eventually calculate the CS + O thermal reaction rates.
   In order to check the accuracy of our calculations, we compare the results of our theoretical calculations for  T$\sim$150$-$400 K with those obtained in the laboratory.}
{Our detailed theoretical study on the CS+O reaction, which is in agreement with the experimental
data obtained at 150-400 K, demonstrates the reliability of our approach. After a careful analysis at lower temperatures, we find that
the rate constant at 10 K is negligible, below 10$^{-15}$ cm$^3$ s$^{-1}$, which is consistent with the extrapolation of experimental data using 
the Arrhenius expression. }
{{ We use the updated chemical network to model the sulfur chemistry
    in Taurus Molecular Cloud 1 (TMC~1) based on molecular abundances determined from
    Gas phase Elemental abundances in Molecular CloudS (GEMS) project observations.
    In our model, we take into account
the expected decrease of the cosmic ray ionization rate, $\zeta_{H_2}$, along the cloud. The
 abundance of CS is still overestimated when assuming the cosmic  value for the sulfur abundance.
}}
  \maketitle

% \end{titlepage}

  \vspace*{1cm}
  \section{Introduction} \label{sec:introduction}

Gas-phase chemistry plays a key role in the star formation process through
critical aspects such as the gas cooling and the ionization fraction.
Molecular filaments can fragment into prestellar cores to a large extent because molecules cool the gas, thus diminishing the thermal
support relative to self-gravity. The ionization fraction controls the coupling of magnetic fields with the gas,
driving the dissipation of turbulence and angular momentum transfer, therefore playing a crucial role in protostellar collapse
and accretion-disk dynamics (see \citealp{Zhao2016, Padovani2013}).
In particular, atomic carbon (C) is the main donor of 
electrons in the cloud surface ($A_V<4$ mag) and, because of its lower ionization potential,
and as long as it is not heavily depleted, sulfur (S) therefore becomes the main electron provider at  higher extinctions.
In the absence of other ionization agents (X-rays, UV photons, J-type shocks),
the ionization fraction is a function of the cosmic-ray ionization rate for H$_2$ molecules, $\zeta _{H_2}$, and of the
elemental gas-phase abundances \citep{McKee1989, Caselli2002}.

Gas phase Elemental abundances in Molecular CloudS (GEMS) is an IRAM 30m Large Program
designed to estimate the S, C, N, and O depletions
and the gas ionization fraction, X(e$^-$)=n$_{e^-}$/n$_H$,
as a function of visual extinction in a selected set of prototypical star-forming filaments
in low-mass (Taurus), intermediate-mass (Perseus), and high-mass (Orion)
star forming regions. Determining sulfur depletion is probably the most challenging goal of this project
because the sulfur chemistry in cold dark clouds remains a puzzling astrochemical problem.
A few sulfur compounds have been detected in diffuse clouds suggesting that the sulfur abundance in these
low-density regions is close to the cosmic value \citep{Neufeld2015}. However, sulfur seems to be depleted in molecular clouds
by a factor of $\sim$3$-$100 compared to its estimated
cosmic abundance \citep{Tieftrunk1994, Ruffle1999, Goicoechea2006, Fuente2019, Vidal2017, Laas2019, Shingledecker-etal:20}.
The depletion of sulfur is observed not only in cold prestellar cores, but 
also in hot cores or corinos, where the icy grain mantles are expected to evaporate \citep{Esplugues2014, Vidal2018},
and in bipolar outflows \citep{Wakelam2005, Holdship2016}.
Chemical models predict that the two main sulfur reservoirs are atomic S
and solid organosulfur compounds, that is, mainly H$_2$S but also the species like OCS
\citep{Vidal2017, Laas2019}, but direct observation of these species remains
difficult. Alternatively, a significant fraction of sulfur can be trapped in
allotropic form, the most abundant of which being S$_4$  \citep{Shingledecker-etal:20},
as also found in laboratory experiments (e.g., \cite{Jimenez2011}); S allotropes can also be
an important sink of sulfur in comets (e.g., \cite{Calmonte-etal:16}).
So far,  only upper limits have been placed on the solid H$_2$S abundance
in the interstellar medium  \citep{Jimenez2011}.
Atomic S has only been detected in some bipolar outflows
using the infrared space telescope Spitzer \citep{Anderson2013}.
Therefore, we need to base our estimation of sulfur elemental
abundance on the observation of minor species and the use of progressively
more complex gas--grain chemical models (see e.g.,  \citealp{Holdship2016, Vidal2017, Navarro2020, Laas2019, Shingledecker-etal:20}). 

\begin{figure}
\includegraphics[width=0.5\textwidth,keepaspectratio]{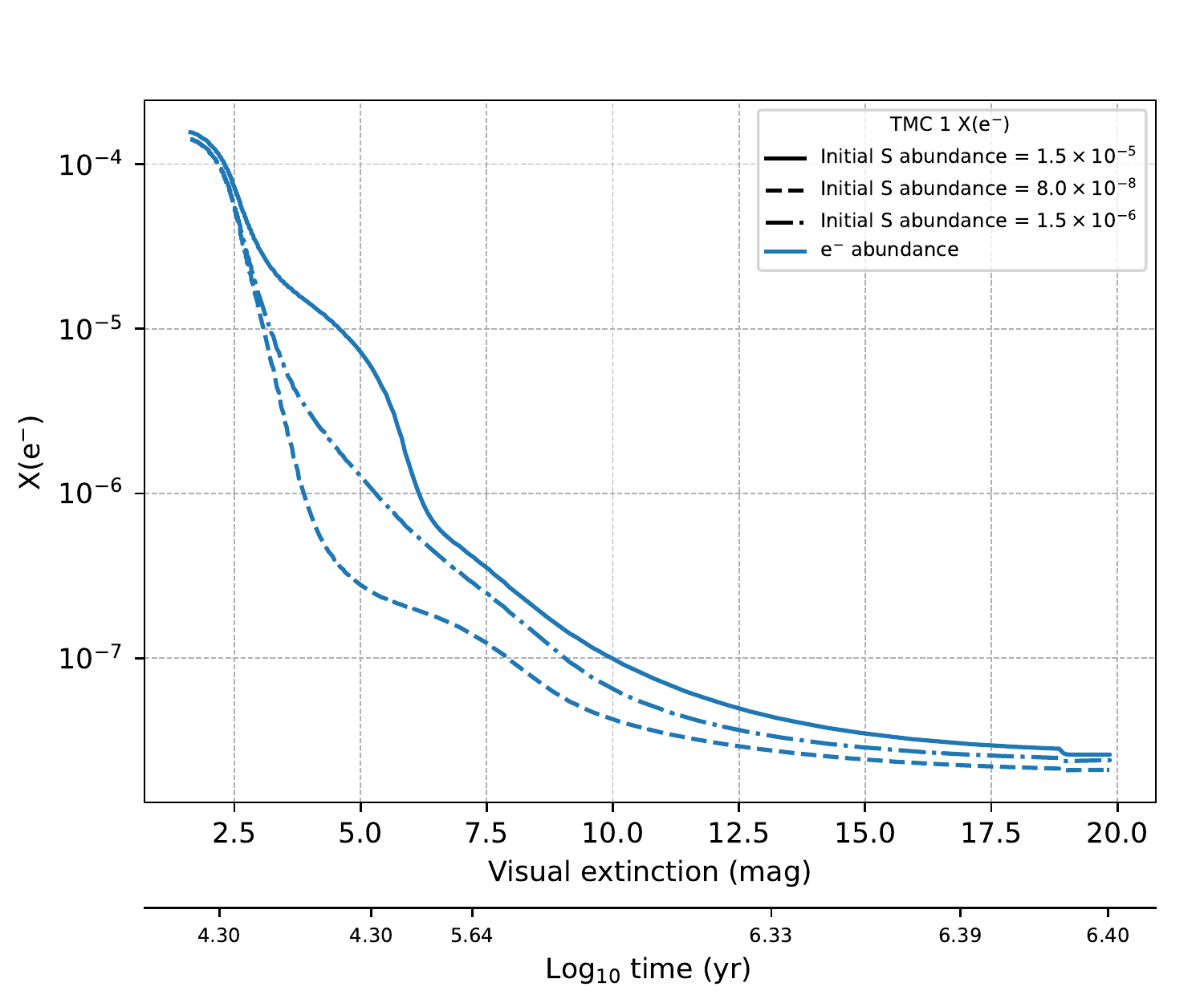}
\caption{Gas ionization fraction, X(e$^-$), as a function of the visual extinction assuming different values of the initial sulfur elemental abundance in TMC~1.
Calculations were performed using the gas-grain chemical code {\scshape Nautilus} \citep{Ruaud2016}, with the physical structure and
the updated chemical network described in \citet{Navarro2020} }
\label{fige}
\end{figure}

The chemistry of sulfur is still poorly understood, with large uncertainties in the gas phase and surface chemical network.
However, a large theoretical and observational effort has been undertaken
in the last five years to understand  sulfur chemistry,
progressively leading to a new paradigm  \citep{Fuente2016, Vidal2017, LeGal2019, Laas2019, Navarro2020, Shingledecker-etal:20}. Based on ab initio
calculations, \citet{Fuente2016} determined the rate of the key reaction S+O$_2$ $\rightarrow$ SO+O at low temperatures.
Using this updated gas-phase chemical network, these latter authors concluded that a moderate S depletion, S/H $\sim(0.6-1.0)\times10^{-6}$,
is necessary to reproduce the high abundances of S-bearing species observed in the dense core Barnard 1b.
This depletion was significantly lower than the usual values adopted in dark clouds \citep{Ruffle1999, Agundez2013}
and some explanations were proposed to explain this overabundance of S-bearing species such as a rapid collapse ($\sim 0.1$ Myr)
that allows most S- and N-bearing species to remain longer in the gas phase, or the interaction of the dense gas with
the compact outflow associated with B1b-S.  The whole gas-phase sulfur chemical network was revised by \citet{Vidal2017}
by looking systematically at the possible reactions between S and S$^+$ with the most abundant species in dense molecular clouds
(CO, CH$_4$, C$_2$H$_2$, and c-C$_3$H$_2$) as well as the potential reactions between sulfur compounds and
the  most abundant reactive species in molecular clouds (C, C$^+$, H, N, O, OH, and CN).
These authors used this new chemical network to interpret previous observations towards
the prototypical dark core TMC1- CP
and found that the best fit to the observations
was obtained when adopting the cosmic sulfur abundance as the initial condition,
and an age of $\sim$1 Myr. Using the same chemical network but with 1D modeling,
\citet{Vastel2018} tried to fit the abundances of 21 S-bearing species towards the starless core L1544.
The authors found that it was impossible to fit all the species
with the same sulfur abundance;  variations of a factor of 100 were found,
and models with initial S/H$\sim8.0\times 10^{-8}$ were those that best fitted the abundances of all 21 species.
New calculations of the SO + OH $\rightarrow$ SO$_2$ + H  reaction rate
reported in \citet{Fuente2019} improved the
description of the SO chemistry at the low temperatures prevailing in dark clouds.
Adopting this new rate and using observations from the GEMS project,
\citet{Fuente2019} derived a sulfur gas-phase abundance of S/H $\sim (0.4-2.2) \times 10^{-6}$ to account for the observations
in the translucent gas (n(H$_2$)$>$10$^4$ cm$^{-3}$) towards the TMC 1 filament. In this paper, the gas-phase PDR Meudon code was used
to fit the observations in the border of this prototypical cloud. Regarding surface chemistry, \citet{Laas2019}
performed an in-depth revision of the surface chemical network in order to incorporate photochemistry,
new results from laboratory experiments, and all the S-bearing molecules detected so far. With this new model,
these latter authors improved the agreement between observations and model predictions assuming the cosmic sulfur abundance. 
Taking into account this more accurate description of the surface chemistry,
\citet{ Shingledecker-etal:20} examined the effects of introducing  cosmic ray-driven radiation chemistry, and  fast nondiffusive bulk reactions for radicals 
and reactive species on the sulfur surface chemistry. These authors showed that these changes have a great impact on the abundances 
of sulfur-bearing species in ice mantles, in particular leading to a reduction in the abundance of solid-phase H$_2$S and HS, and a significant increase 
in the abundances of OCS, SO$_2$, and allotropes of sulfur such as S$_8$.

GEMS provides a complete (the most abundant species) and spatially resolved
(measurements at different visual extinctions within the same cloud down to $A_V\sim3$ mag)
database of sulfur-bearing species, which allows extensive comparison with models
to describe the progressive sulfur depletion along the cloud,
and eventually allows us to estimate the initial S/H. \citet{Navarro2020} carried out a detailed physical
and chemical modeling of the cores TMC1-CP, TMC1-C, and Barnard 1b in an attempt to explain the observed CS,
SO, and H$_2$S observations, which are the most abundant gas-phase S-bearing species present in these clouds.
To do so, \citet{Navarro2020} used the chemical model  {\scshape Nautilus},
which was recently updated by \citet{LeGal2019}
to include the  most recent observations, reaction coefficient rates,
and S-chemical pathways \citep{Fuente2016, Fuente2017, Vidal2017, Fuente2019}, and then by themselves to
incorporate the new surface reaction network by \citet{Laas2019}.  
Finally,  \citet{Navarro2020} took into account chemical desorption using the prescriptions of \citet{Minissale2016} for
bare and ice-coated grains. One of the results of that paper was that the authors were unable to fit the CS,
SO, and H$_2$S abundances simultaneously. While the SO and H$_2$S abundances
were well fitted with their chemical model assuming the cosmic sulfur elemental abundances,
the CS abundance was overestimated by a factor of more than ten.
This lack of accordance prevents us from determining a reliable value
for the initial S/H abundance which remains with an uncertainty of a factor of more than ten,
varying between S/H $\sim 10^{-6}$ and 1.5$\times 10^{-5}$.
 \citet{Navarro2020} recall that different initial S/H abundances would lead to a different gas ionization fraction.
In Fig.~\ref{fige}, we predict X(e$^-$) using 
the chemical model described by these latter authors and different initial values of S/H.
It should be noted that  X(e$^-$) varies by more
than a factor of ten for $A_V$$<$ 10~mag, depending of the initial value of S/H,
which becomes a key parameter to model the fragmentation
of molecular filaments to form dense cores.

\section{CS chemical network}
CS is among the most abundant gas phase S-bearing molecules in dark clouds. It is easily observable with several transitions in the
millimeter wavelength range, and has a simple rotational spectrum with well-known collisional coefficients \citep{Denis2018, Lique2006a}. 
Therefore, it has been largely used as a density and column density tracer in the interstellar medium  in our Galaxy and external 
galaxies (see, e.g., \citealp{Snell1984, Lapinov1998, Kim2020, Martin2005, Bayet2009, Bayet2015}). 
Moreover, CS is the only S-bearing molecule routinely detected in protoplanetary disks and is therefore the main tracer of sulfur abundance in
the primordial material to form planets \citep{Agundez2018,LeGal2019}. 
An understanding of  CS chemistry is essential for correct interpretation of the observations 
from all astrophysical environments.
Unfortunately, chemical models do a poor job at accounting for these observations, usually predicting CS abundances  much larger than
 those observed \citep{Gratier-etal:16, Vidal2017}.

 The chemistry of CS in interstellar clouds is closely correlated with that of HCS$^+$
 and involves reactions
 that have never been studied experimentally, leading
 to large uncertainties.  For very young molecular clouds, where the ionization fraction
 from the diffuse period is still very large, sulfur is essentially
 in atomic ionized form and controls the chemistry of sulfur (e.g., \cite{Goicoechea2006}).
 CS is then produced essentially from the electronic dissociative recombination (DR)
 of HCS$^+$, HCS$^+$ being produced by the S$^+$ + CH$_2$ and CS$^+$ + H$_2$ reactions, and
 CS$^+$ being produced by the ion-neutral reactions S$^+$ + CH and S$^+$ + C$_2$.
 For the more advanced stages of dense clouds, which probably more closely correspond
to the clouds observed in the GEMS project,
 the ionic fraction is much lower and the sulfur is mainly in neutral atomic form
 (the reactions of ionized atomic sulfur are not negligible but play a secondary role).
 Under these conditions, although the DR of HCS$^+$ still produces CS,
 HCS$^+$ is also mostly formed from CS (either directly by the CS + H$_3^+$ reaction,
 or indirectly by CS + H$^+$ $\rightarrow$ H + CS$^+$
 followed by CS$^+$ + H$_2$ $\rightarrow$ HCS$^+$ + H) and not from S$^+$ reactions. 
 In that case, CS is produced by neutral reactions,
 mainly S + CH and S + C$_2$, with secondary contributions by H + HCS, S + CH$_2$ , and C + SO.
 The overestimation of CS in the models versus the observations
 could come from an underestimation of the rates of consumption reactions
 (mainly CS + H$^+$ and CS + H$_3^+$).
 However, this seems unlikely because even if there are no measurements,
 the rates used are those resulting from the capture theory and thus close to the maximum theoretical rates. The CS overestimation could also come from overestimation of the production rates from neutral reactions such as S + CH and S + C$_2$, or from missing consumption reactions of CS. For the latter case, \citet{Vidal2017} suggested that a high rate for the reaction of CS with the abundant atomic oxygen, O + CS,  will  decrease the overproduction of CS without 
heavily affecting the abundance of the S-bearing molecules, except for the chemically related HCS$^+$. 
This possibility motivated the present study to better quantify the O + CS reaction rate.

Chemical models use the CS + O reaction rate constants measured by \cite{Lilenfeld-Richardson:77}  in the 150-300 K interval
considerably higher than the typical T$_k$$\sim$10 K of dark clouds,which are then extrapolated to low temperatures using the Arrhenius expression.
The extrapolation to lower temperatures is always questionable and experimental measurement and/or theoretical calculations are needed
to confirm these values.   \cite{Gonzalez-etal:96} ran theoretical simulations
by calculating the potential energy surface (PES) for the ground and first excited states
and obtained reaction rate constants under several transition state theory (TST) approaches.
However, the values found by these latter authors at 150$-$300 K were considerably lower than those seen in experimental measurements, casting doubts
about the accuracy of the calculated rates.
It is therefore necessary to improve the theoretical simulations to predict reasonable reaction rates
at the lower temperatures prevailing in the interstellar medium (ISM).

This study is devoted to the theoretical determination of the CS+O reaction rate.
The {\it ab initio} calculations performed to produce the lower potential energy surfaces (PESs)
are described in Section 3. These PESs are then used to study the reaction dynamics,
using several methods (classical, quantum, and semiclassical) to derive the reaction rates.
Finally, we test the role of the new reaction rates on realistic chemical models of cold dark clouds.
%The complete quantum description is computationally very demanding to produce thermal 
%rate constants at low temperatures.

\section{Potential energy surfaces}\label{sec:pes}

Calculation of the PES is a mandatory step for any dynamical study of a chemical reaction. The reaction

\begin{eqnarray}\label{CS+O-reaction}
  {\rm CS}(X^1\Sigma^+) + {\rm O}(^3P)  \rightarrow {\rm CO} (X^1\Sigma^+) + {\rm S}(^3P)
\end{eqnarray}
involves open-shell atoms in reactants and products, presenting three degenerate electronic
states at long distances (neglecting spin-orbit), correlating to  P states of the oxygen or sulfur
atoms. At long distances, the energies of these three states are dominated by the dipole-quadrupole
interactions \citep{Buckingham:67}. However, at short distances there are 
excited electronic states correlating to ${\rm CS}(a ^3\Pi) + {\rm O}(^3P)$ \citep{Gonzalez-etal:96}, which cross with the
lower electronic manifold, giving rise to the formation of the ${\rm CO} (X^1\Sigma^+) + {\rm S}(^3P)$ products.
These crossings give rise to small barriers whose height strongly depends on the electronic basis and the method chosen to 
describe the electronic correlation, as noted by \cite{Gonzalez-etal:96}. 

In this work accurate {\it ab initio} calculations are performed using the internally contracted multi-reference configuration interaction (ic-MRCI)  method
\citep{Werner-Knowles:88,Werner-Knowles:88b} including the Davidson correction (icMRCI+Q) \citep{Davidson:75}.
In these calculations, the molecular orbitals are optimized using a state-averaged complete active space self-consistent field (SA-CASSCF) method,
with an active space of 14 orbitals (11 and 3  of  $a'$ and $a''$ symmetry, respectively). One $^3A'$ and two $^3A''$ electronic states are calculated 
and simultaneously optimized. In all these calculations the aug-cc-pVTZ basis set is used \citep{Dunning:89}.
For the ic-MRCI calculations, seven orbitals are kept doubly occupied, giving rise to $\approx$ $30 \times 10^6$ $(6500 \times 10^6)$
contracted (uncontracted) configurations. All {\sl ab initio} calculations were performed with the MOLPRO suite of programs \citep{MOLPRO-WIREs}.

The analytical representation of the adiabatic PESs is done in three parts:
\begin{enumerate}
  \item For short to intermediate distances,   a three-dimensional cubic
spline method is used  with the DB3INK/DB3VAL subroutines based on the method of \cite{Boor:78} and distributed by GAMS \citep{gams-www}.
A dense grid is calculated, composed of 20$\times$14$\times$19 points in the intervals defined
in bond coordinates as: R$_{CO}$ ($\lbrack 0.9, 10\rbrack$ \AA), R$_{CS}$  ($\lbrack 1, 7\rbrack$ \AA), and $\Theta_{OCS}$ ($\lbrack 0, \pi\rbrack$),
respectively.

\item At long distances (R$_{CO}>8$ \AA), dipole-quadrupole long-range interactions
  are considered using the expressions defined by \cite{Zeimen-etal:03} in reactant
  Jacobi coordinates.  The V(R$_{CS}$) obtained at R$_{CO}$=100 \AA\ is fitted using the diatomic terms
  of \cite{Aguado-Paniagua:92}. The CS electric dipole is fitted as a function
  of the R$_{CS}$ distance, and the O($^3P$) quadrupole is calculated as energy derivatives using different
  homogeneous electric fields \citep{MOLPRO-WIREs}. The long-range behavior 
  is checked  by doing  ic-MRCI calculations for distances $R$ longer than 10 \AA, with 
  $R$ being the distance between the CS center of mass and the oxygen atom.

\item In order to guarantee a continuous behavior  between the previous two regions,
  points calculated with the long-range expression are added at R$_{CO}$=7, 8, and 9 \AA,
  and a damping function among the two regions is centered at 5 \AA. 

\end{enumerate}

%%%%%%%%%%%%%%%%%%%%%%%%%%%%%%%%%%%%%%%%%%%%%
\begin{figure}[t]
\begin{center}
  \includegraphics[width=7.cm]{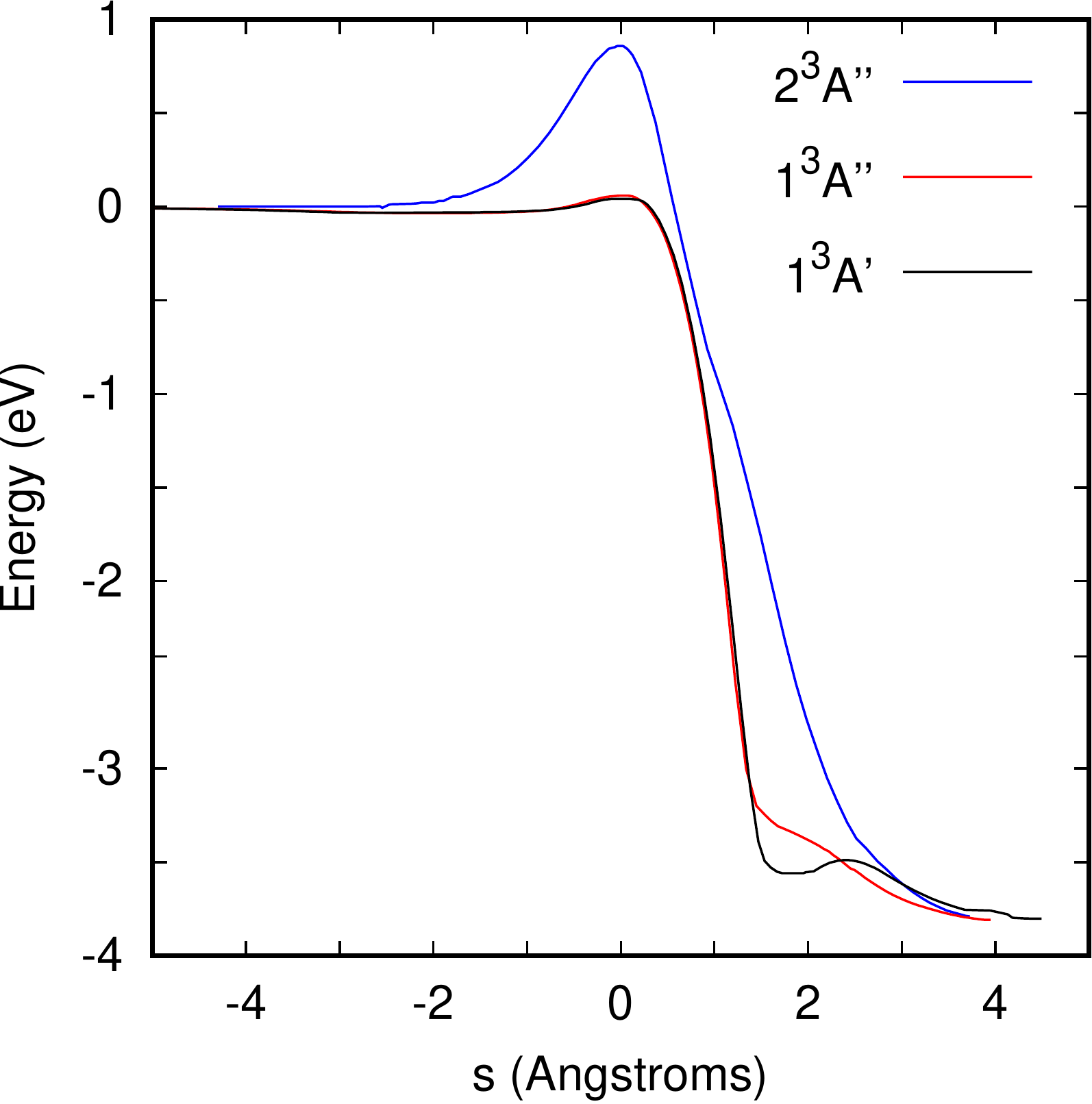}
  
  \caption{\label{energy-diagrams-CS+O-reaction}{ Three lower adiabatic potential energy surfaces
      as a function
        of the intrinsic reaction coordinate (IRC)
       for  the $
      {\rm CS}(X^1\Sigma^+) + {\rm O}(^3P)  \rightarrow {\rm CO} (X^1\Sigma^+) + {\rm S}(^3P)$ reaction.
  }}
%\vspace*{0.1cm}
\end{center}
\end{figure}
%%%%%%%%%%%%%%%%%%%%%%%%%%%%%%%%%%%%%%%%%%%%%

The minimum energy path for the reaction 
is shown in Fig.~\ref{energy-diagrams-CS+O-reaction}
for the three adiabatic states (1 $^3A'$ and 2 $^3A''$).
The reaction is exothermic by $\approx$ 3.9 eV, in agreement with
the value of 3.93 eV reported by \cite{Gonzalez-etal:96}. When zero-point energy (ZPE)
is taken into account, the exothermicity reduces to 3.59 eV in rather good
agreement with the experimental value of  3.64 eV \citep{Lilenfeld-Richardson:77}.
The energy barriers obtained in this work are 0.043, 0.058, and 0.888 eV for the 1$^3$A', 1$^3$A'', and
2$^3A''$ states, respectively. These values are lower than those obtained by \cite{Gonzalez-etal:96},
probably because the electronic correlation introduced by ic-MRCI is higher than the PUMP4 method.

%%%%%%%%%%%%%%%%%%%%%%%%%%%%%%%%%%%%%%%%%%%%%
\begin{figure}[t]
\begin{center}
 \includegraphics[width=9.cm]{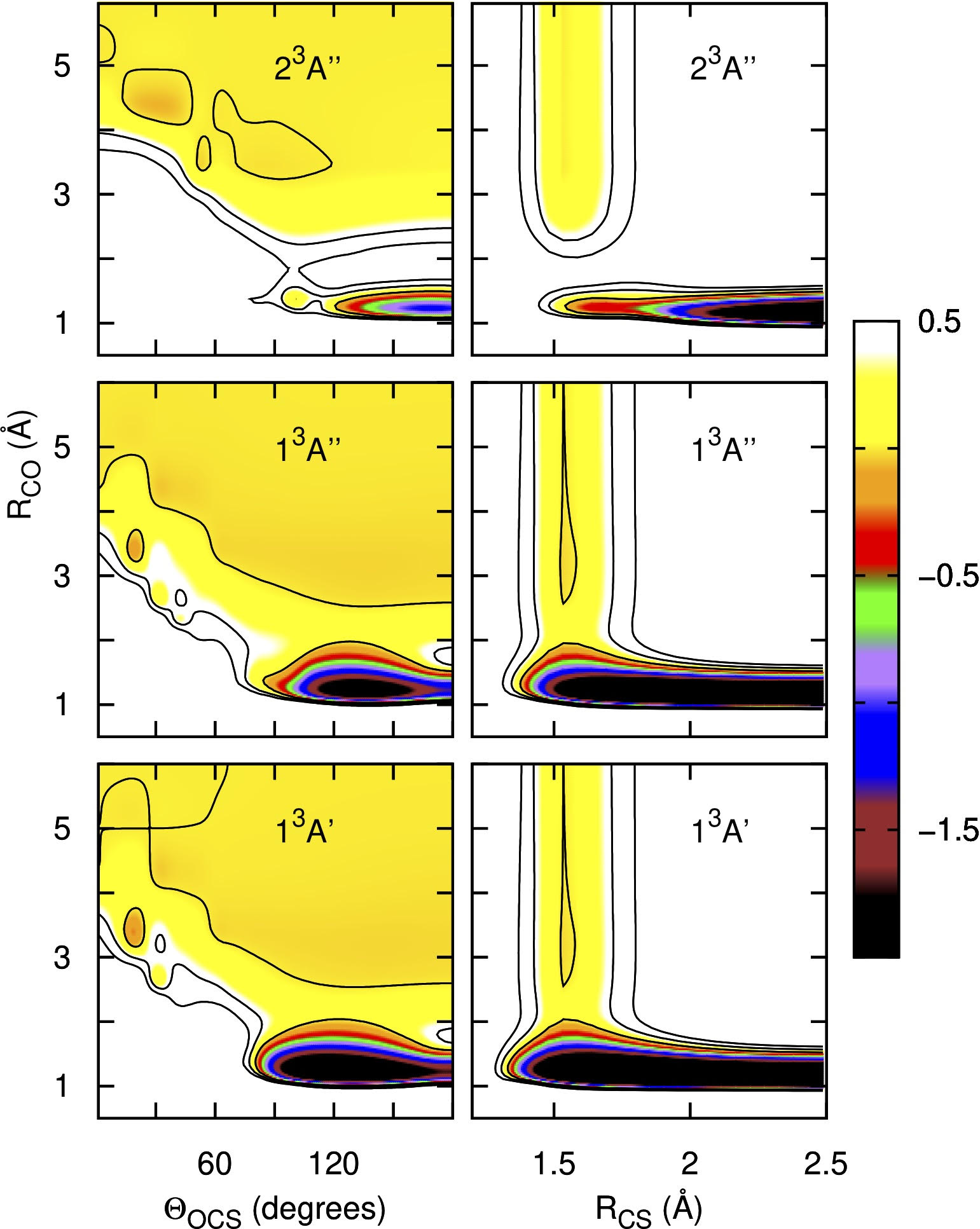}
  
  \caption{\label{PES-cuts}{Contour plots of the PES for the three electronic
      states obtained at the equilibrium R$_{CS}$= 1.535 \AA\ as a function of R$_{CO}$ and the OCS angle
      (left panels)
      and at an OCS angle of 120$^o$ as a function of R$_{CO}$ and R$_{CS}$ distances.
      Energies are in eV, and the contour lines are at 0, 0.5, and 1 eV.
  }}
%\vspace*{0.1cm}
\end{center}
\end{figure}
%%%%%%%%%%%%%%%%%%%%%%%%%%%%%%%%%%%%%%%%%%%%%

The main features of the present PESs are very similar to those discussed by \cite{Gonzalez-etal:96},
represented in the contour plots shown in Fig.~\ref{PES-cuts}. The reaction barriers
are located in the entrance channel, at nearly the equilibrium distance of CS,
and at R$_{CO}\approx$2.25~\AA~ for the ground electronic state.
In addition, the angular cone of acceptance is also reduced as $R$ distance becomes
closer: the saddle point is located at OCS angle,
$\Theta_{OCS} \approx$ 120$^o$, and the interval is reduced to $\lbrack 80^o,160^o\rbrack$. According
to the Polanyi rules, the early barrier suggests that translational energy will enhance the reactivity.
The reduction of the angular cone of acceptance is expected to introduce some restrictions, as  discussed below in the reaction dynamics section.

\section{Reaction dynamics}\label{sec:dyn}

The thermal reaction rate can be defined as
\begin{eqnarray}\label{themal-rate} 
  K(T)= \sum_{vje} w_{vje}(T)\, K_{vje}(T) \quad{\rm with}\quad w_{vje}= {e^{-E_{vje}/k_BT} \over \sum_{v'j'e'} e^{-E_{v'j'e'}/k_BT} }
,\end{eqnarray}
where the sum is over all vibrational, rotational, and electronic states of the reactants,
CS(X$^1\Sigma^+$, v j) + O($^3P$),
of energy $E_{vje}$.
Here, $ K_{vje}(T)$ are the initial state selected rate constants, which correspond to
the Boltzmann average over the translational energy of the reaction cross-section,
\begin{eqnarray}\label{state-selected-rate} 
  K_{vje}(T) = \sqrt{ {8\over \pi \mu (k_BT)^3}}  \int dE\, E\, \sigma_{vje}(E) e^{-E/k_BT}.
\end{eqnarray}

The cross-section is obtained under the partial wave summation over
the total angular momentum, $J$, as
\begin{eqnarray}\label{cross-section}
  \sigma_{vje}(E) = {\pi \over (2j+1)\,k^2_{vj}(E)} \sum_{J\Omega} (2J+1)\, P^J_{vje\Omega} (E),
\end{eqnarray}
where $k_{vj}= \sqrt{2\mu E}/\hbar$ (with $\mu$ being the CS + O reduced mass), $\Omega$ is
the helicity, that is, the projection of ${\bf J}$ and ${\bf j}$ angular momenta on the z-axis of the body-fixed frame,
and $P^J_{vje\Omega} (E)$ is the reaction probability for a particular initial state of
the reactants, which depends on collision energy $E$. This quantity can be calculated
with different methods: exact and approximate, quantum and classical. Below
we start by determining the accuracy of each of them for $J$=0.

\begin{figure}[t]
\begin{center}
 \includegraphics[width=8.cm]{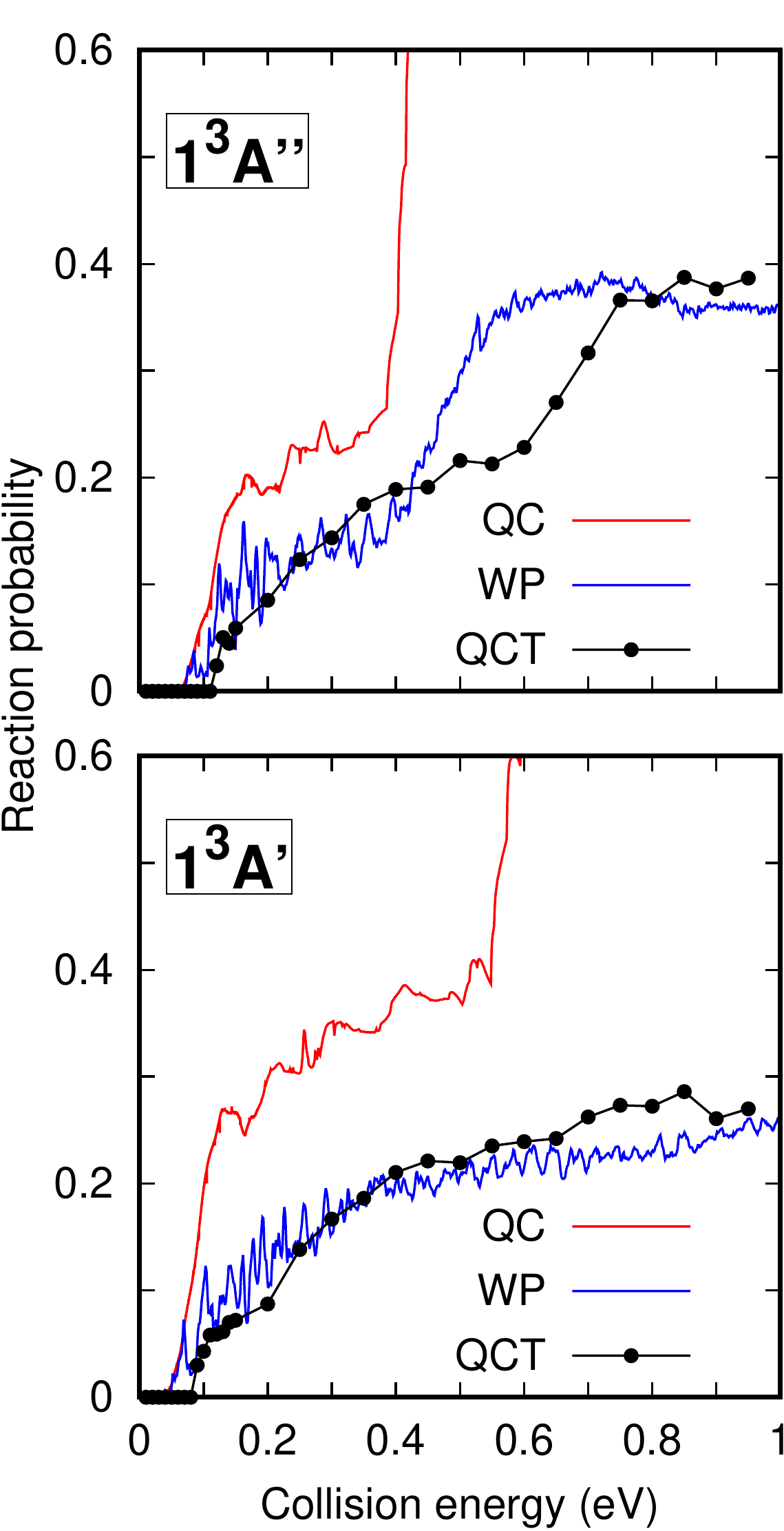}
  
 \caption{\label{probJ0}{CS+O $\rightarrow$ CO + S reaction probabilities vs. collision energy 
     for $J=0$ in the 1 $^3A'$ and 1 $^3A''$ electronic states using three different methods described
     in the text: the QC, the WP,  and the QCT
     methods.
  }}
%\vspace*{0.1cm}
\end{center}
\end{figure}

The reaction is very exothermic, but it presents a reaction barrier. It can be assumed
that all the flux that passes over this barrier yields products,  considerably reducing
the computational effort. This can be done using the quantum capture approach \citep{Clary-Henshaw:87},
in which the time-independent  close coupled equations (TICCEs) are solved in the entrance channel, similarly
to what it is done in inelastic collisions, but subject to capture conditions, that is, to outgoing
complex conditions at $R<R_c$ for those channels for which E$> V_{vje\Omega}(R_c)$, with $R_c = 2$\AA\ 
being the capture distance. Thus the TICCEs are integrated from R= 2 \AA\  to 30 \AA\  in the rovibrational
states composed of CS(v=0,1,2) and 50 rotational states for total angular momentum $J$=0.
This is done separately for each electronic state, 1$^3A'$ and 1$^3A''$
using the ZTICC code \citep{Gomez-Carrasco-etal:20}. The capture
probabilities are compared with quantum wave packet (WP) results in Fig.~\ref{probJ0}. These
calculations were performed with the MADWAVE3 code \citep{Zanchet-etal:09b} and the parameters
used are listed in Table ~\ref{wvp-parameters}. The WP method is considered numerically exact, but as discussed below, it is very computationally demanding.
 \begin{table}[h]
 \caption{\label{wvp-parameters}
   Parameters used in the wave packet calculations in reactant Jacobi coordinates:
   $r_{min} \leq r\leq r_{max}$ is the CS internuclear distance, $R_{min} \leq R\leq R_{max}$ is the distance between CS center of mass and the oxygen
   atom, $0 \leq \gamma \leq \pi$ is the angle between ${\vec r}$ and ${\vec R}$ vectors. The initial wave packet
   is described in $R$ by a Gaussian centered at $R=R_0$, and at a translational energy of
   $E=E_0$, and width $\Delta E $. The total reaction probability is obtained by analyzing the total flux
   at $r=r_\infty$
}
 \begin{center}
 \begin{tabular}{|cc|}
 \hline 
 $r_{min}$, $r_{max}=$  0.1, 10 \AA & $N_r$=512 \\
 $r_{abs}$=  5 \AA & \\
$R_{min}$, $R_{max}=$   0.001, 18\AA & $N_R$=1024  \\
 $R_{abs}$=  11 \AA  &\\
$N_\gamma$ = 240 & in $[0,\pi]$  \\
$R_0$  = 9 \AA & $E_0,\Delta E$= 0.4,0.2 eV\\
$r_\infty$ = 4 \AA &   \\
 \hline
 \end{tabular}
 \end{center}
 \end{table}

 Clearly the quantum capture (QC) method overestimates the reaction probability.
 Near the reaction threshold,
 the QC and WP results are in rather good agreement,
 showing a common threshold at 0.04 and 0.06 eV for
 1$^3A'$ and 1$^3A''$, respectively. However, above the threshold energy,
 the QC method gives
a much larger reaction
probability than the WP method. This is clear evidence
that not all the flux arriving at distances
$R$ shorter than $R_c$ go on to form CO + S products,
and this situation increases
with increasing collision energy.

%%%%%%%%%%%%%%%%%%%%%%%%%%%%%%%%%%%%%%%%%%%%%
%%%%%%%%%%%%%%%%%%%%%%%%%%%%%%%%%%%%%%%%%%%%%

As the reaction involves rather heavy atoms, it may be
expected that quantum effects do not play an important role.
The quasi-classical trajectory (QCT) method is then an interesting
alternative to simplify the computationally demanding quantum WP calculations.
The comparison for $J$=0 in Fig.~\ref{probJ0} reveals rather
good agreement, except at the threshold. The QCT method is not
able to describe the first peaks appearing in the WP reaction probabilities, which
can be attributed to tunneling.

To better quantify the adequacy of the QCT method to describe this reaction,
we calculated the total cross-section with the QCT and WP methods. In order to limit the highly
demanding WP calculations for high $J$,
we performed the centrifugal sudden approximation (CSA); \citep{Pack:74,McGuire-Kouri:74},
in which only one helicity $\Omega$ is included. Also, we calculated the reaction
probability  for $J$=0, 50, 100, 150 and 180, and the reaction probabilities
for the remaining $J$s are obtained using an interpolation based on the $J$-shifting approximation
\citep{Aguado-etal:97,Zanchet-etal:13}. The comparison between the WP-CS and QCT calculations is shown
in Fig.~\ref{cross-section}, and they show reasonably good agreement below 0.2 eV. However, for
higher energies, the QCT cross-sections are in general higher than the WP-CS ones, and the differences
are larger for 1$^3A'$ than for 1$^3A''$. At these higher energies one would
expect better agreement between classical and quantum methods, similar
to that obtained for $J=0$. The larger difference can be attributed to the CS approximation
made to obtain the cross-section in the case of the quantum WP-CS method.
In order to check this, for the 1$^3$A' state and
$J$ = 50, 100, and 150 we included  more helicities on the reaction
probabilities,  $\Omega=$ 0,1, 2, 3, 4, and 5. These new calculations, labeled
`WP' in Fig.~\ref{cross-section}, are larger than the WP-CS calculations, and very close
 to the QCT calculations up to 0.3 eV. Above this energy, more helicities $\Omega$
are needed to converge the reaction probabilities of $J$>100. However, these calculations
are extremely demanding.

%%%%%%%%%%%%%%%%%%%%%%%%%%%%%%%%%%%%%%%%%%%%%
\begin{figure}[t]
\begin{center}
 \includegraphics[width=8.cm]{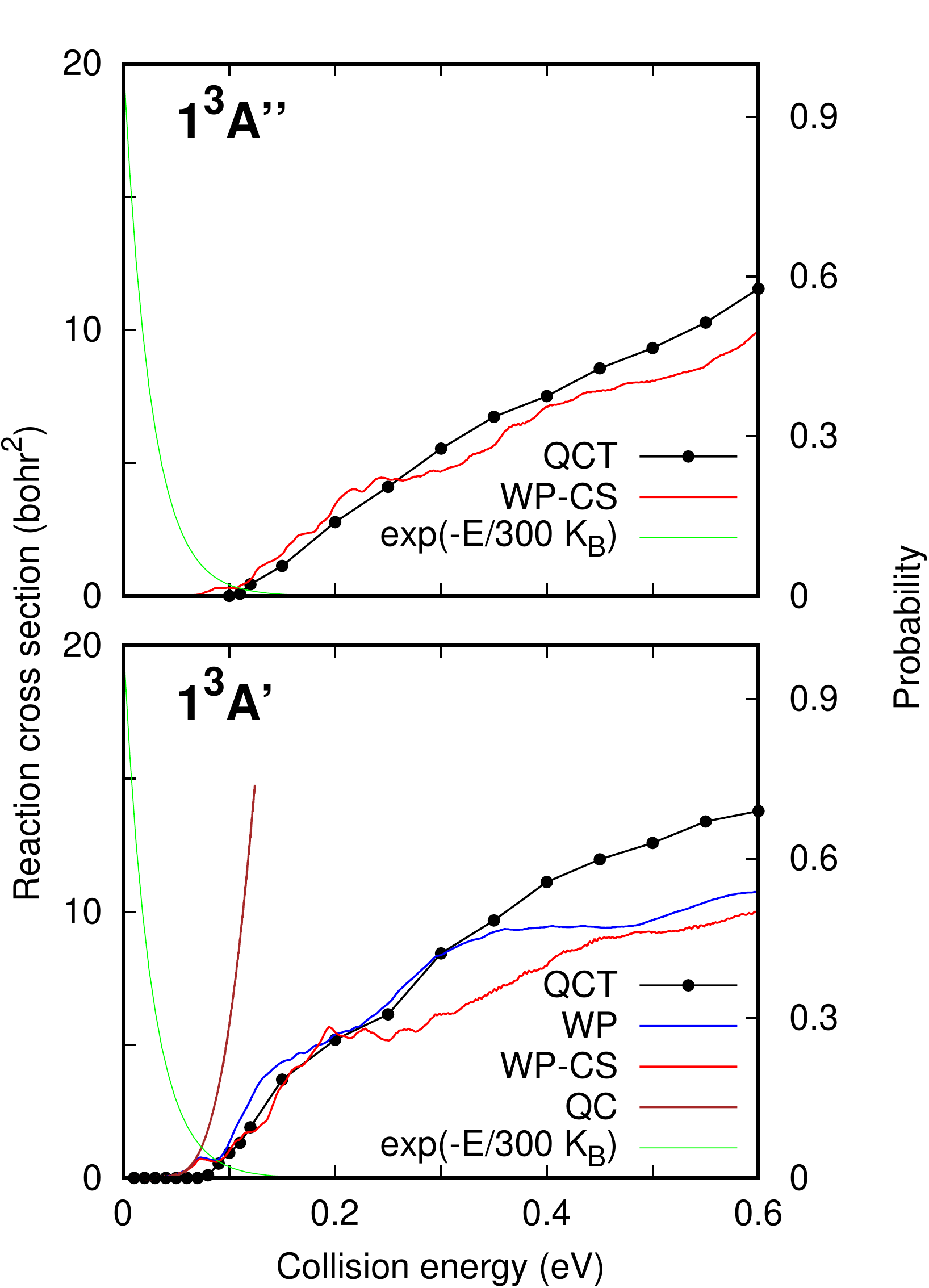}
  
 \caption{\label{cross-section}{CS+O $\rightarrow$ CO + S reaction cross-section  (in Bohr$^2$) vs. collision energy
     for  1 $^3A'$ and 1 $^3A''$ electronic states using the quantum wave packet within the CS approach
     (WP-CSA)  and the QCT
     methods. `QC' labels the results obtained with the QC method.
     The energy distribution of a Boltzmann distribution for a temperature of 300 K is also shown in green.
     For the  1 $^3A'$, WP labels the wave-packet calculations performed including $\Omega$=0,1,2,3,4 and 5.
  }}
%\vspace*{0.1cm}
\end{center}
\end{figure}
%%%%%%%%%%%%%%%%%%%%%%%%%%%%%%%%%%%%%%%%%%%%%

In Fig.~\ref{cross-section} the probability arising for a Boltzmann distribution at 300 K is also displayed, showing that
only collision energy below 0.12 eV contributes for temperatures below 300 K. Below 0.12 eV, QCT results
are lower than the quantum wave packet values. This indicates that it is important to include quantum effects
near the threshold. WP methods require individual calculations for each initial state, and many rotational
states  have to be considered (which contribute significantly below 0.12 eV) because of the low rotational constant of CS.
This makes
the use of the WP method  in evaluating the thermal rate constants for this reaction very computationally demanding,
 and some alternative method should be used.
 {
   The QC results for energies below 0.07 eV are in very good agreement with the exact WP calculations.
  However, QC results clearly overestimate the reaction cross-section for higher energies. Nevertheless,
  the Boltzmann energy distribution  corresponds to collision energies below 0.07 eV
  for temperatures below 150 K and therefore the QC results can be considered
  to be nearly exact in this low temperature range.
}

Ring polymer molecular dynamics (RPMD) is a semiclassical method based on path integral methods
that include quantum effects such as zero-point energy and tunneling proposed by \cite{Craig-Manolopoulos:04}.
RPMD has been successfully applied to calculate reaction rate constants \citep{Craig-Manolopoulos:05,Craig-Manolopoulos:05b,Suleimanov-etal:11}
as recently reviewed by \cite{Suleimanov-etal:16}. Here we apply a direct version of this method recently applied
to reactions of polyatomic molecules at low temperature  \citep{Suleimanov-etal:18,delMazo-Sevillano-etal:19,Bulut-etal:19}
and implemented in the code dRPMD.

%%%%%%%%%%%%%%%%%%%%%%%%%%%%%%%%%%%%%%%%%%%%%
\begin{figure}[t]
\begin{center}
 \includegraphics[width=8.cm]{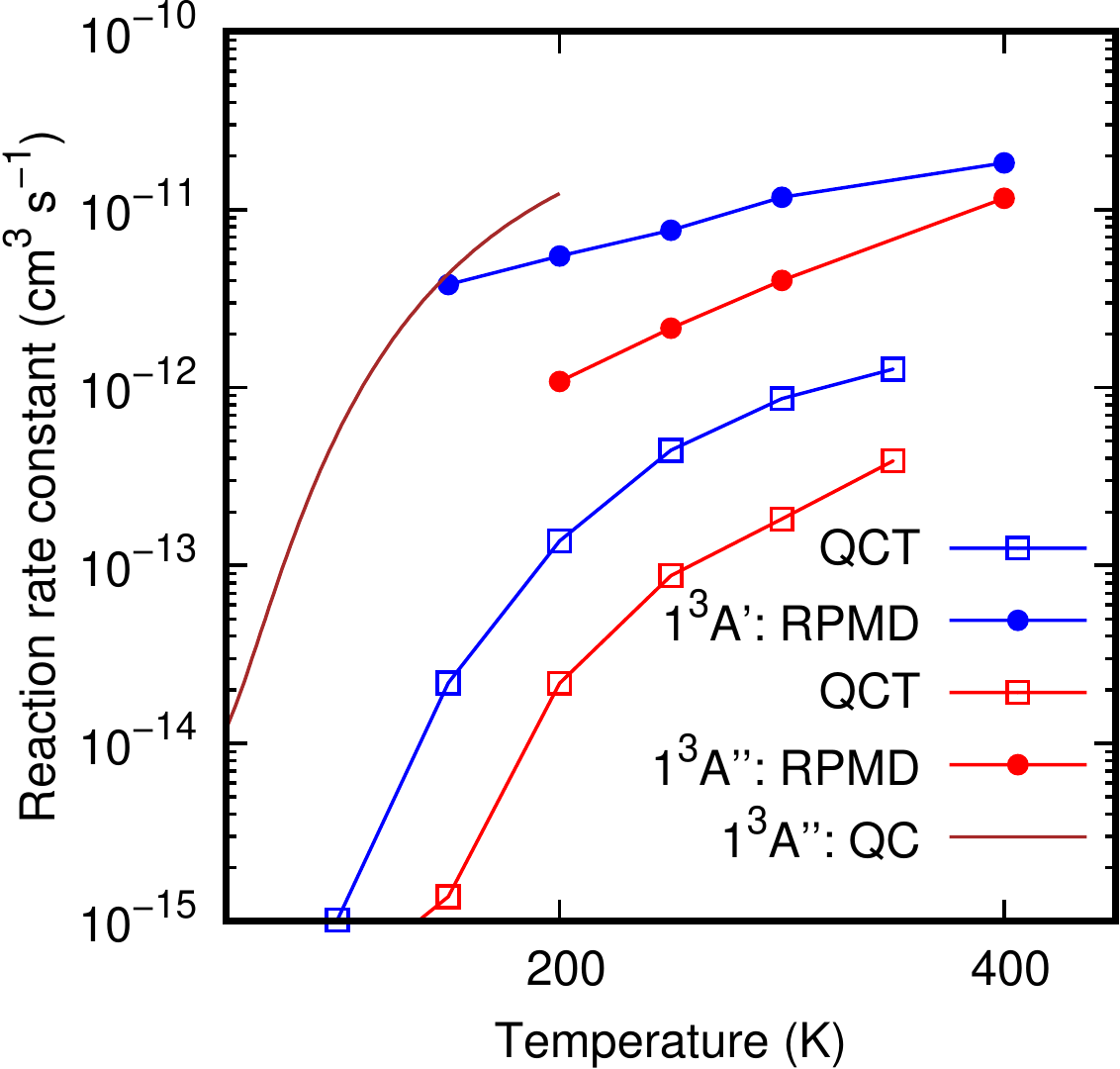}
 \caption{\label{RPMDvsQCTrates}{CS+O $\rightarrow$ CO + S reaction rate constants obtained
     with RPMD (full circles) and QCT (open squares) for the 1$^3A'$ (blue) and 1$^3A''$ (red) electronic states.
  }}
%\vspace*{0.1cm}
\end{center}
\end{figure}
%%%%%%%%%%%%%%%%%%%%%%%%%%%%%%%%%%%%%%%%%%%%%

RPMD, QCT, and QC results are compared in Fig.~\ref{RPMDvsQCTrates} for the 1 $^3A'$ and 1 $^3A''$ electronic states.
The QCT calculations consist of more than 10$^5$ trajectories
per temperature (for low temperatures, more than 10$^6$ trajectories were needed to get convergence).
RPMD results are based on 10$^4$ trajectories using a variable number of beads (64 for 300 K, 128 for 150 K, etc).
RPMD rate constants are always about ten times larger than the QCT ones. This is explained by the difference
found in the cross-section obtained with quantum WP and QCT methods at energies below 0.12 eV. RPMD includes
quantum effects and the results show that it is more accurate than the QCT.
It is important to stress here that, according to QC calculations, the reaction probability at low energies
increases with the initial rotational state of the CS reagent. The QCT and RPMD rate calculations
include this effect by considering the rotational temperature, and this produces an amplification of the difference
between QCT and RPMD rate constants.
In both cases, many trajectories have been
run for temperatures below 100 K, but no reactive ones were found. This indicates that the reaction rate constant
below 100 K is very small. 
{
  The QC results at 150 K are very close to the RPMD rate constant. For 200 K, however, QC results are
  considerably larger. This result is expected as QC overestimates
  the reaction probabilities above 0.07 eV. However, for temperatures below 150 K
  it is expected to be a rather good upper limit of the reaction rate constant.
}

\begin{figure}[t]
\begin{center}
 \includegraphics[width=8.cm]{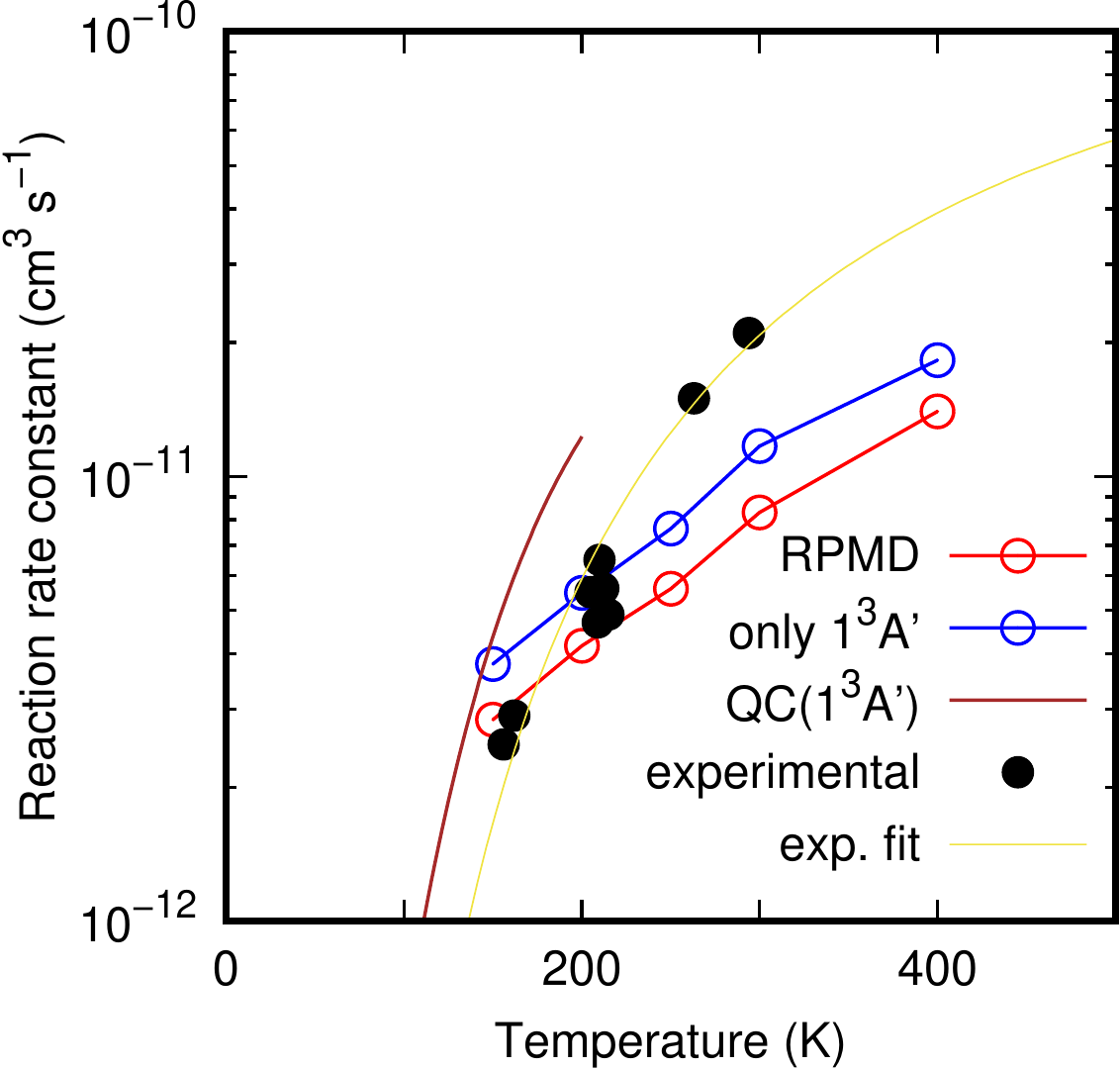}
  
 \caption{\label{RPMDvEXPrates}{Comparison of the calculated thermal rate constant
     for the CS+O $\rightarrow$ CO + S reaction including spin-orbit
     splitting and the experimental measurements of \cite{Lilenfeld-Richardson:77}. The rate constant obtained for  1$^3A'$  is included for
     discussion. The experimental results are fit to k=A e$^{-C/T}$, with A=2.6 10$^{-10}$ cm$^3$s$^{-1}$ and C=757.7 K=0.065 eV.
  }}
%\vspace*{0.1cm}
\end{center}
\end{figure}

The thermal rate constant is finally obtained by an average over the spin-orbit electronic states
of O($^3P$)  as
\begin{eqnarray}
  k(T)= {   3 k^{1^3A'}(T) +   k^{1^3A''}(T)  \left( 2 + e^{-227.71/T}\right)
    \over 5 + 3 e^{-227.71/T} +  e^{-326.98/T} },
\end{eqnarray}
where an adiabatic approximation has been made for the spin-orbit states,
and $k^{2^3A''}=0$ . The results are compared with the experimental
    results of \cite{Lilenfeld-Richardson:77} in Fig.~\ref{RPMDvEXPrates}.
    The results of the calculations presented here are close to the experimental values for T=150-200 K, becoming  a factor of between two and three smaller
    at 300 K.
    According to the fit to the Arrhenius law shown in Fig.~\ref{RPMDvEXPrates}
    the activation energy is $\approx$ 0.065 eV, while the potential energy
    barriers obtained here are lower, namely 0.043 and 0.058 for
    the 1$^3A'$ and 1$^3A''$ states, respectively. Furthermore, the rate constant obtained
    for the 1 $^3A'$ state alone is very close to the experimental
    value, changing the slope of the rate constant versus temperature.
    The disagreement at 300 K is attributed to inaccuracies of the 1,2$^3A''$ excited electronic states.
    Also, as RPMD includes quantum effects such as tunneling and zero-point energy effects, we may conclude
    that the rate constant decreases with temperature,  following an Arrhenius law.
{ This Arrehenius-like behavior is found in the QC results below 150 K, confirming
  the behavior in the fitted rate constant to the experimental values (note that
  the rate constant for the 1$^3A''$ is lower).
  }
    Therefore, we may conclude that at the temperatures relevant in
    dense molecular clouds, T$_k \sim$10 K, the CS + O reaction rate constant is negligible, below 10$^{-15}$ cm$^3$s$^{-1}$.

%%%%%%%%%%%%%%%%%%%%%%%%%%%%%%%%%%%%%%%%%%%%%
%%%%%%%%%%%%%%%%%%%%%%%%%%%%%%%%%%%%%%%%%%%%%

\section{Astrophysical implications}

\begin{figure}
%\begin{center}
 \includegraphics[width=1\linewidth]{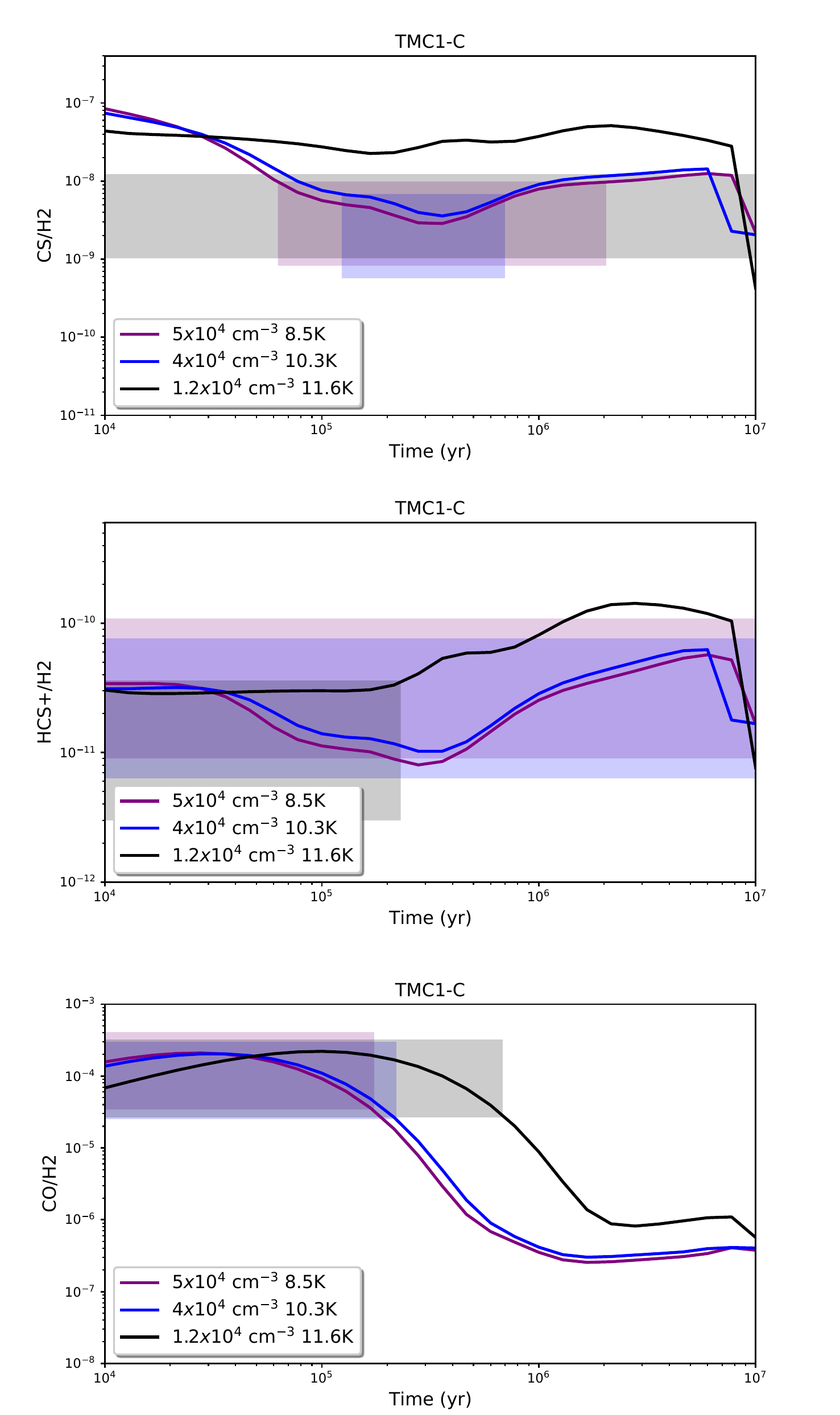}
 \caption{\label{TMC1C_allratios_zeta1e-16} Predicted abundances of gas-phase CS, HCS$^+$,
   and CO with respect to H$_2$ as a function of time.
   The curves correspond to the different physical conditions observed in the TMC1-C source
   with  $\zeta_{H_2}$ = $10^{-16}$~s$^{-1}$. Each curve (density) corresponds
   to a different position of Table~3
   in \citet{Fuente2019}. Colored boxes represent the agreement with the observations.
   The abundance of S with respect to H is depleted by a factor of 20 relative to the cosmic value.}
%\vspace*{0.1cm}
%\end{center}
\end{figure}
\begin{figure}
%\begin{center}
 \includegraphics[width=1\linewidth]{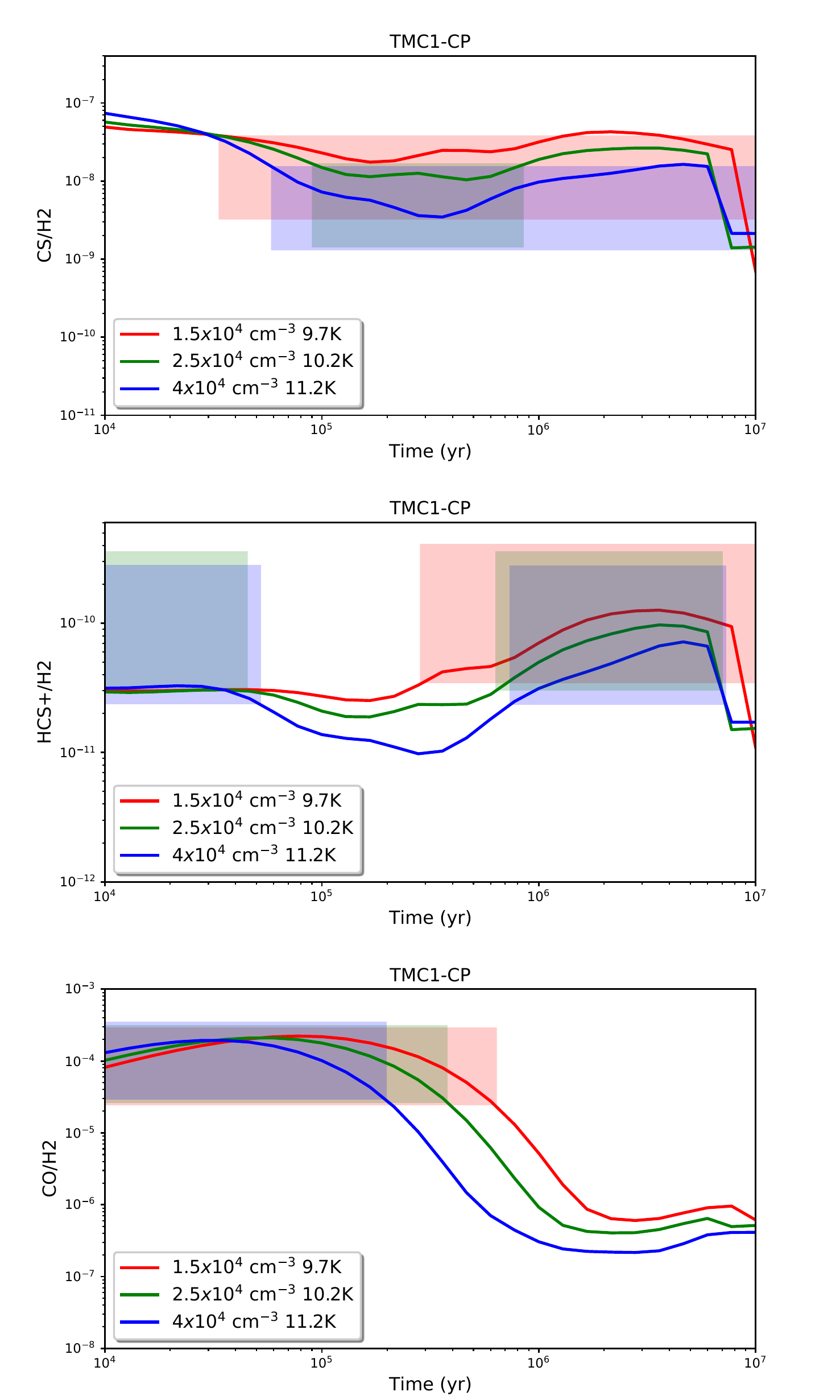}
 \caption{\label{TMC1CP_allratios_zeta1e-16} Same as Fig.~\ref{TMC1C_allratios_zeta1e-16} but for TMC1-CP.}
%\vspace*{0.1cm}
%\end{center}
\end{figure}
\begin{figure}
%\begin{center}
 \includegraphics[width=1\linewidth]{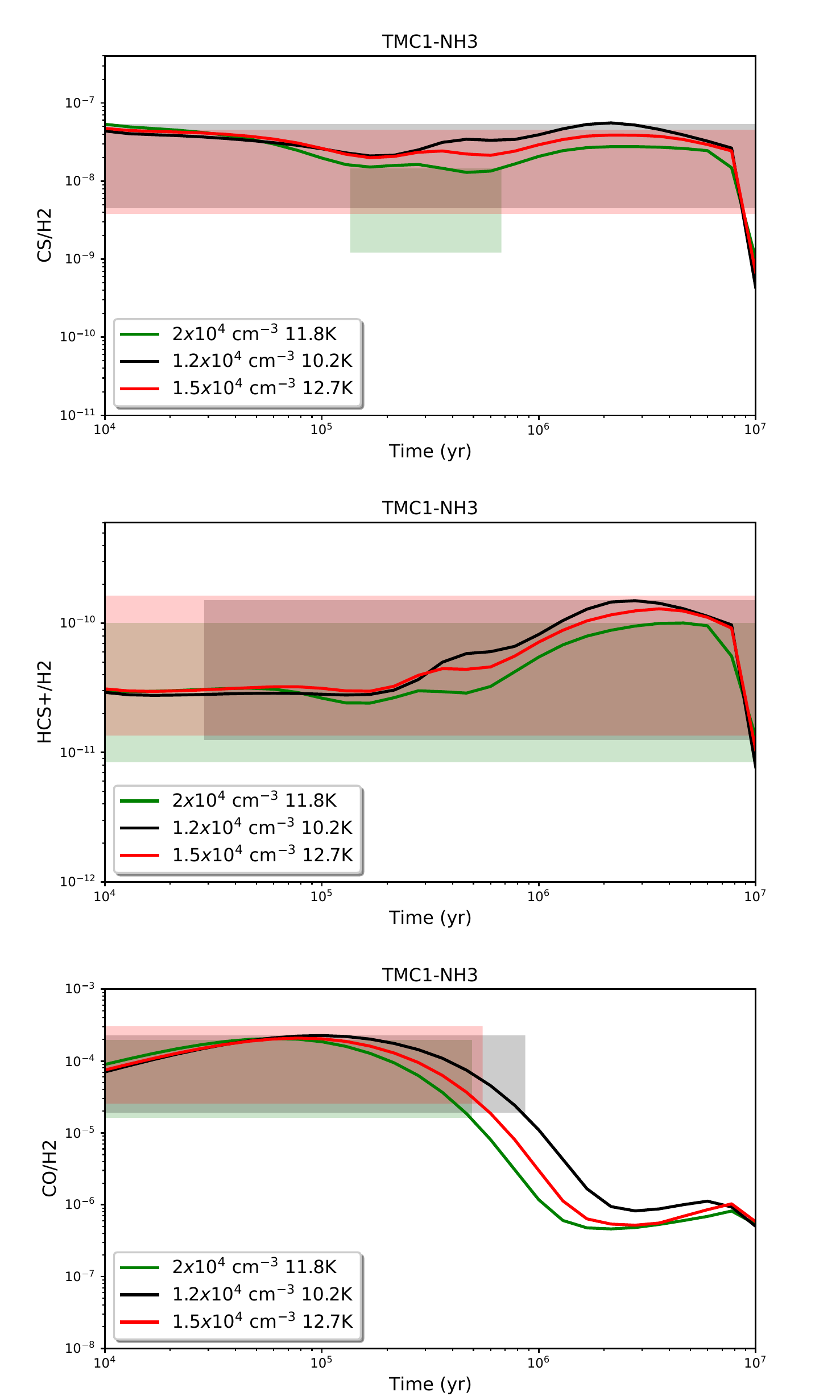}
 \caption{\label{TMC1NH3_allratios_zeta1e-16} Same as Fig.~\ref{TMC1C_allratios_zeta1e-16} but for TMC1-NH3.}
%\vspace*{0.1cm}
%\end{center}
\end{figure}

\begin{figure}
%\begin{center}
 \includegraphics[width=1\linewidth]{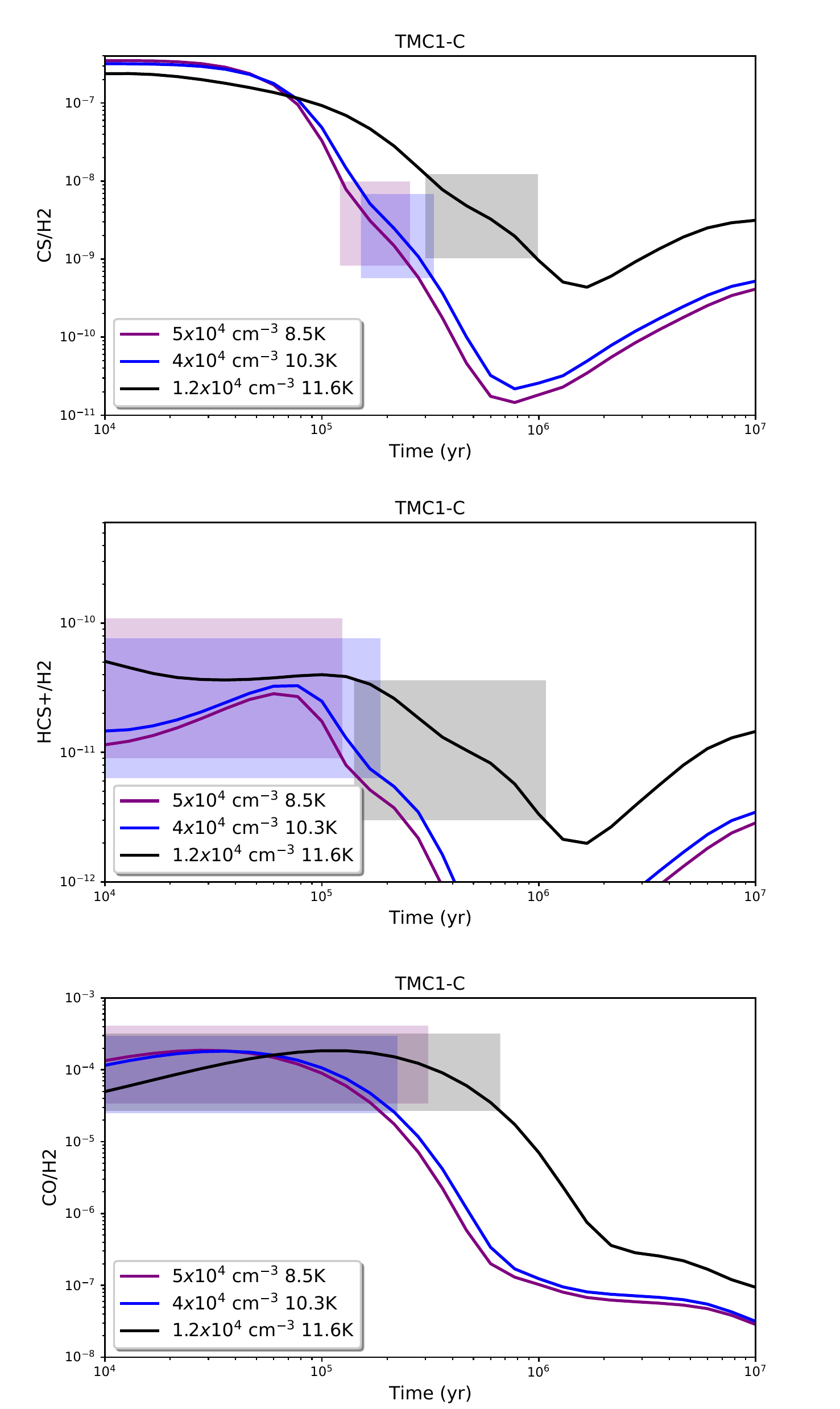}
 \caption{\label{TMC1C_allratios} Same as in Fig. 8  but with  $\zeta_{H_2}$
   calculated for each visual extinction (see text and \citealp{Indriolo-etal:12,Neufeld-etal:17}).
}
%\vspace*{0.1cm}
%\end{center}
\end{figure}
\begin{figure}
%\begin{center}
 \includegraphics[width=1\linewidth]{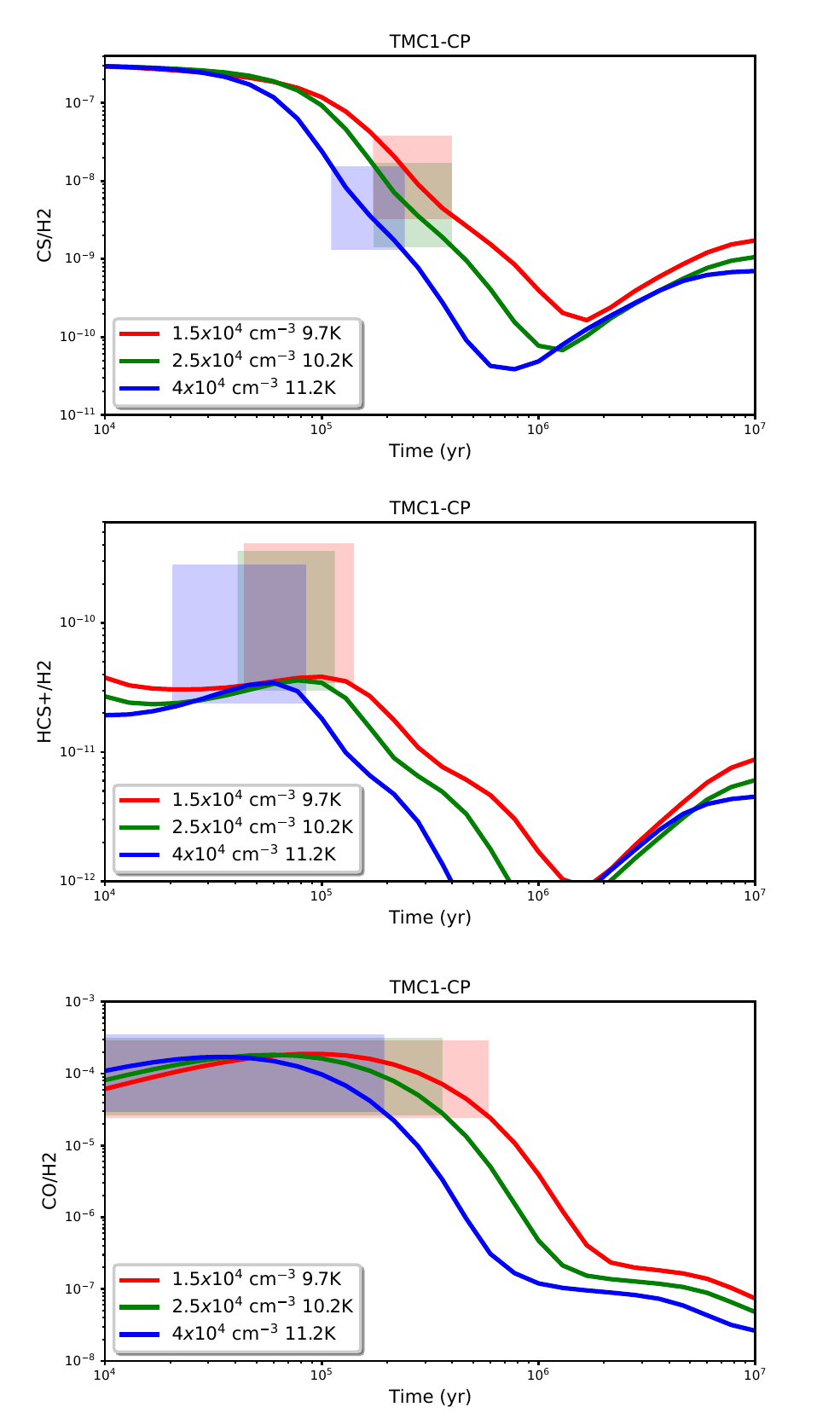}
 \caption{\label{TMC1CP_allratios} Same as in Fig. \ref{TMC1CP_allratios_zeta1e-16},
   but with  $\zeta_{H_2}$ calculated for each visual extinction.}
%\vspace*{0.1cm}
%\end{center}
\end{figure}
\begin{figure}
%\begin{center}
 \includegraphics[width=1\linewidth]{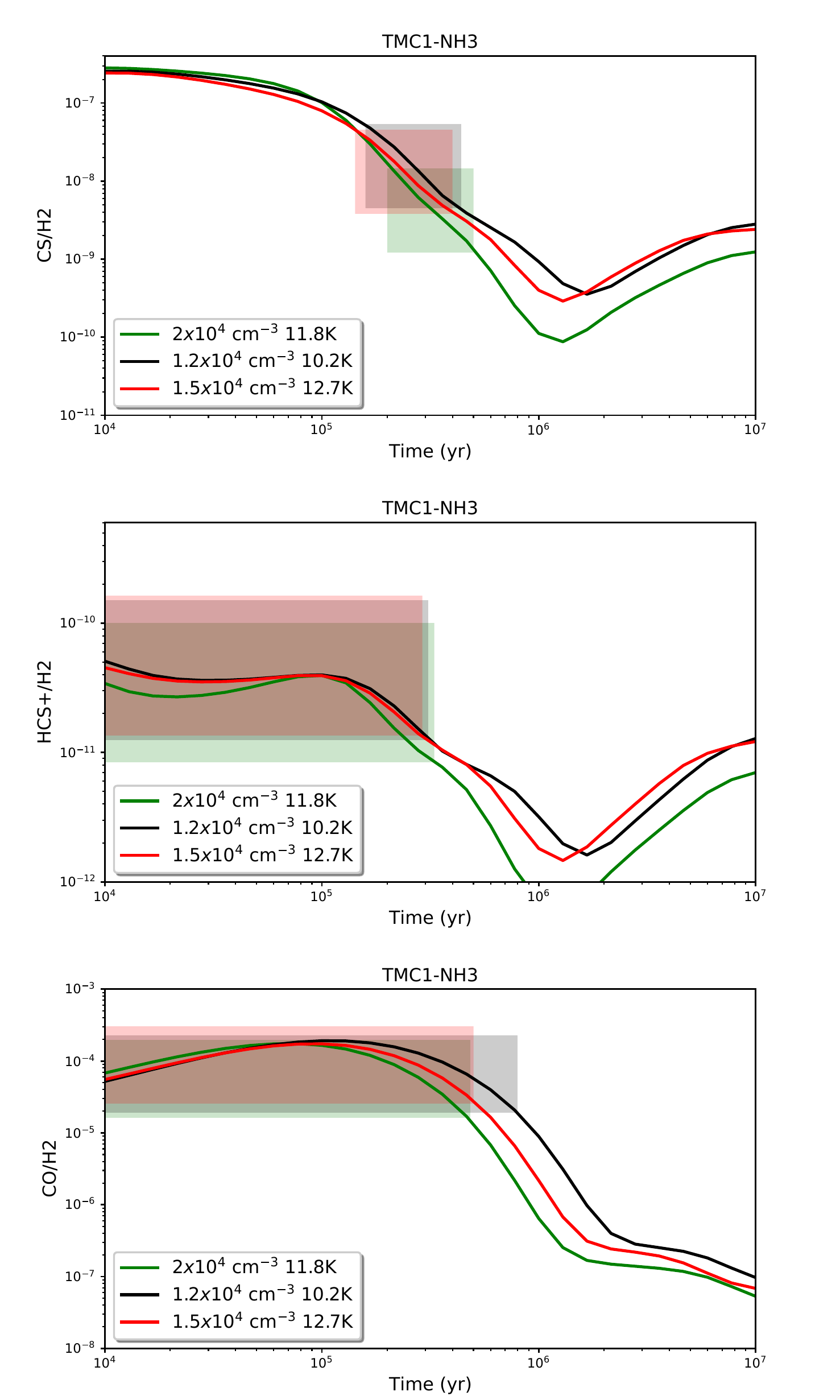}
 \caption{\label{TMC1NH3_allratios} Same as in Fig.\ref{TMC1NH3_allratios_zeta1e-16}
   but with  $\zeta_{H_2}$ calculated for each visual extinction.}
%\vspace*{0.1cm}
%\end{center}
\end{figure}

Here we present a detailed theoretical study of the CS+O reaction, confirming the experimental
data obtained at 150-400 K, and after a careful analysis at lower temperatures we find that
the rate constant at 10K is negligible, below 10$^{-15}$ cm$^3$ s$^{-1}$.
Given the low value of the rate constant of the CS + O reaction at low temperature,
this reaction does not seem to be able to explain the calculated overabundance of CS given by dense cloud models.
A  CS + O reaction rate close to 1$\times 10^{-10}$  cm$^3$s$^{-1}$ at 10~K,
five orders of magnitude higher than our limit, would be needed to account for the observed CS abundances if no ad hoc depletion of sulfur is assumed.
In addition to the O + CS reaction, the chemical network for the destruction reactions of CS seems to us complete
and relatively precise. The overestimation of CS does not seem to be due to an underestimation of the CS destruction
reactions. Another hypothesis for the cause of this overestimation, previously put forward in the section above (CS chemical network),
could be an overestimation of the CS production reactions. For a typical chemical evolution of the clouds
corresponding to the observations, CS is mainly produced by neutral reactions, that is, mainly S + CH and S + C$_2$.
The rates for these two reactions in the model are close to those given by the capture theory,
which may overestimate the value. 
A decrease in these rates would lead to a decrease in the production of CS because, despite their importance, the fluxes of 
these reactions are smaller than the fluxes of the  S + H$_3^+$, S + OH, S + CH$_3$ reactions considering 
the CH, C$_2$, H$_3^+$, OH, and CH$_3$ abundances given by the model
(and for some of them by the observations) considering the physical conditions of the studied dark clouds.
To our knowledge, there are no experimental data for or theoretical studies of the S + CH and S + C$_2$ reactions.
Indeed, there is very little information on S + radical reactions in general.
\citet{Flores-etal:01}  performed a theoretical
study of the S + C$_2$H reaction leading to a very high rate constant at low temperature,
similar to the O + C$_2$H one \citep{Georgievskii2011}.
Therefore, as the O + CH reaction is rapid at room temperature characteristic of a barrierless reaction \citep{Messing1980},
we may expect similar behavior for the S + CH reaction and a high rate constant at low temperature.
An overestimation  of the S + CH and S + C$_2$ reactions required to reproduce the CS abundances by more than a factor of at ten seems unlikely.
Nevertheless, it is clear that theoretical and experimental studies are needed to better characterize
S + radical reactions.

An additional problem comes from the fact that
if the abundance of CS decreases, the abundance of HCS$^{+}$
would also decrease in typical dense clouds, because in this case HCS$^{+}$
is mainly produced from CS. Furthermore, as the measured abundances of HCS$^{+}$ are
significantly higher than the modeled abundances, the decrease of CS will accentuate
the disagreement. Either there is an unknown direct production
(not from CS) of HCS$^{+}$ or the destruction of HCS$^{+}$ is overestimated.
As the DR of HCS$^{+}$ is by far the main loss of HCS$^{+}$,
a smaller value of the rate constant for this DR will increase the HCS$^{+}$ abundance.
This DR has been experimentally studied by \citet{Montaigne2005}
and there are no specific reasons to question this value. Nevertheless,
there is only an experimental value and it should be noted that the DR
of HCNH$^{+}$ \citep{Adam1991, Semaniak2001, McLain2009}
and N$_2$H$^{+}$ \citep{Shapko2020} vary greatly from one measurement to another. 
New experimental measurements of the DR of HCS$^+$ would be desirable to confirm 
the currently used value.

{
In order to evaluate the impact of our calculated CS + O reaction rate, we modeled 
the GEMS data along the dense clouds TMC 1-C, TMC 1-CP, and TMC 1-NH3. These data were recently presented
and modeled by \citet{Fuente2019} and \citet{Navarro2020}. Here, we resumed this modeling 
focusing on {CO, CS, and HCS$^{+}$ }.  For this modeling we used the updated 
network from \cite{Vidal2017} and the same temperatures, densities, incident UV flux, and visual extinction ($A_V$) 
of each observed region as those used by  \citet{Navarro2020}. For the cosmic-ray molecular hydrogen ionization rate, $\zeta_{H_2}$, we use  
the fixed value  equal to $10^{-16}$s$^{-1}$ as  determined by \citealp{Fuente2019}.
 The results of these simulations are presented  in Figs.~\ref{TMC1C_allratios_zeta1e-16} to \ref{TMC1NH3_allratios}  for CS, HCS$^{+}$, and CO. 
%  The abundance of CO has been calculated
%  from observations of the C$^{18}$O 1$\rightarrow$0 line, and assuming N(CO)/N(C$^{18}$O)=600 (see \citealp{Fuente2019}).
In these figures, the colored boxes represent 
the period of time in which model predictions agree with observed abundance ratios at each position.
An uncertainty of a factor of two is assumed for the observed abundances, which translates into an uncertainty of a 
factor of four in the molecular abundance ratios.
The abundance of CO  makes it possible to accurately constrain
  the maximum age of the clouds since CO is rapidly depleted under the physical conditions of these clouds.
%As can be seen in Figs.~\ref{TMC1C_allratios_zeta1e-16},~\ref{TMC1CP_allratios_zeta1e-16}
%  and ~\ref{TMC1NH3_allratios_zeta1e-16}, the cosmic-ray ionization rate has a strong effect on the abundances
%  of CS and HCS$^{+}$, and little effect on CO. For $\zeta_{H_2}$ = 10$^{-16}$~s$^{-1}$,
  The CS profile is flat and the only way to obtain a good agreement between the observations
  and the model is to strongly deplete the sulfur elemental abundance (by a factor of 20 in the curves shown in Figs.~\ref{TMC1C_allratios_zeta1e-16} to \ref{TMC1NH3_allratios}).
  In this case, the agreement for CS and HCS$^{+}$ can only be considered as `satisfactory',
   as HCS$^{+}$ is underestimated for TMC 1-CP, while CO is also fairly well modeled.
However, with such a sulfur depletion factor,  the H$_2$S abundance would remain underestimated by a factor of more than ten
(see \citealp{Navarro2020}), thus challenging our comprehension of the sulfur chemistry.}

{
A crucial point in the modeling of sulfur compounds, in addition to the sulfur depletion factor,
is the specific dependency of their abundances on the cosmic-ray ionization rate (see e.g., \citealp{Fuente2016}). It is
known that the cosmic ray flux decreases with A$_V$ following a law that is
dependent on the local conditions \citep{Padovani2009, Padovani2013, Ivlev2018,Padovani-etal:18}. 
Thus far, we have used a fixed value of   $\zeta_{H_2}$  in our simulations.
One may postulate that the disagreement between chemical predictions and observations is due to the adopted
fixed value for $\zeta_{H_2}$.  In order to evaluate this effect, 
we repeated the simulations assuming  $\zeta_{H_2}$ to change with $A_V$. In particular we
assumed a different value of  $\zeta_{H_2}$ for each A$_V$ following the fit shown in Fig. 6 of \citealp{Neufeld-etal:17}).}

 \begin{equation}
 log_{10}(\zeta_{H_2}) = -1.05 \times log_{10} (A_V) - 15.69
.\end{equation}
 
 {
 This expression gives values of  $\zeta_{H_2}$  $\sim$ $10^{-17}$~s$^{-1}$ for an $A_V$ of 13 mag
  and $\sim 4\times 10^{-17}$~s$^{-1}$ for an $A_V$ of 5 mag. These values are significantly lower than the value
  previously adopted ($\zeta_{H_2}$ = 10$^{-16}$~s$^{-1}$). 
  Figures~\ref{TMC1C_allratios},~\ref{TMC1CP_allratios}, and ~\ref{TMC1NH3_allratios} show
  model predictions using the new values of $\zeta_{H_2}$.
  Interestingly, the value of $\zeta_{H_2}$ has a great impact on the CS and HCS$^+$ abundances, but its impact is
  negligible for CO.  With these new values, the CS profiles are much more stepped, with a shape 
 similar to that of CO. In this case, one can always find a cloud age that allows reproduction of CS regardless of the sulfur depletion factor.
 However, the obtained ages are only compatible with 
 CO abundances when we assume a low value for the elemental sulfur abundance. 
 More specifically, the ages required to account for the observed CS abundances are too large for CO, which
 would be strongly depleted on the grain surfaces unless we assume that sulfur elemental abundance is depleted by a factor of $\sim$20.
 Figures~\ref{TMC1C_allratios_zeta1e-16},~\ref{TMC1CP_allratios_zeta1e-16},
  and ~\ref{TMC1NH3_allratios_zeta1e-16} show the comparison between models and observations assuming that the elemental sulfur abundance is
  depleted by a factor 20 with the observations. The agreement is good for CS but
  in this case it is more difficult to reproduce the abundance of HCS$^{+}$ for ages where CS is reproduced. In addition,
  we still do not reproduce the H$_2$S abundance \citep{Navarro2020}. We therefore conclude  that decreasing the cosmic ray 
  flux with A$_V$ does not  help to find a better agreement between chemical models and observations. }
 
Our study on the rate of the O + CS reaction removes one of the hypotheses
for the overestimation of CS in the models versus the previous observations. 
{ The new analysis of GEMS observations  using an updated chemical network shows the importance of the cosmic-ray
ionization rate on the predicted abundances of sulfur-bearing species in cold dark clouds. However, we are not able to reproduce observations by 
decreasing $\zeta_{H_2}$ with the visual extinction. Further  observational and theoretical research is
needed. }
From the observational point of view, it would be desirable to  complete the molecular database with important sulfur-bearing species
other than CS, HCS$^{+}$, and H$_2$S, in particular C$_2$S, C$_3$S, OCS, and H$_2$CS 
to better constrain the value of the cosmic-ray ionization rate  and its coupling with the sulfur depletion factor.
 From the theoretical point of view, there is still room for
 significant improvement. Despite recent reviews on the chemistry of sulfur \citep{Fuente2017, Vidal2017, Laas2019, Fuente2019, Navarro2020,
 Shingledecker-etal:20}, the rates and branching ratios of sulfur chemistry reactions are too poorly known, which prevents the models from being really predictive.
A substantial theoretical and experimental effort on the rates of neutral atomic sulfur reactions,
on the branching ratios of S$^{+}$ reactions and on HCS$^+$ DR rate  is needed if we hope to better understand 
the chemistry of sulfur in the interstellar medium.   

\section{Conclusions}

The CS+O reaction has been proposed as a relevant CS destruction mechanism at low temperatures.
Its reaction rate has been experimentally measured at temperatures of 150$-$400~K,
but the extrapolation to lower temperatures is uncertain. In this study, we
calculated the CS+O reaction rate at temperatures $<$150 K which are prevailing in 
 cold dark clouds.
 We performed {\it ab initio} calculations to produce the lower potential energy surfaces (PESs) of the CS +O system. These PESs are  used to study the reaction dynamics,
 using several classical, quantum, and semiclassical  methods
 to eventually calculate the CS + O thermal reaction rates.
 In order to check the accuracy of our calculations,
 we compared the results  with those obtained in the laboratory over the  T$\sim$150$-$400 K range.
 We present a detailed theoretical study of the CS+O reaction,
 the results of which are in agreement with experimental
data, verifying the reliability of our approach.
After careful analysis at lower temperatures we find that
the rate constant at 10 K is negligible,
below 10$^{-15}$ cm$^3$ s$^{-1}$, consistent with the extrapolation of experimental data using the Arrhenius expression.

We modeled observations of CS and HCS$^+$ using  an updated chemical network. We obtain a good fit of the CS, HCS$^+$, and SO 
abundances assuming a sulfur depletion by a factor of 20 and different chemical ages for each position within the cloud. Still, the H$_2$S abundance 
would remain underestimated by a factor of more than ten unless  we assume no sulfur depletion (S/H=1.5$\times$10$^{-5}$). 
{ We also investigated the effect of the decrease of  $\zeta_{H_2}$ with A$_V$ on the abundances of S-bearing species.
Still, we need to adopt a sulfur depletion by a factor of 20 if we want to fit the abundances of CO, CS, and HCS$^+$ using the same chemical age.  This
high depletion would lead to underestimation of the H$_2$S abundance. Therefore, further theoretical and observational research is needed to understand
the sulfur chemistry.}
In spite of recent efforts to complete and
update sulfur chemistry \citep{Fuente2017, Vidal2017, Laas2019, Fuente2019, Navarro2020, Shingledecker-etal:20}, there are still many uncertainties in the chemical network.
A substantial theoretical and experimental effort on the rates of neutral atomic sulfur reactions,
on the branching ratios of S$^{+}$ reactions, and on the HCS$^+$ DR rate  is needed if we hope to better understand 
the chemistry of sulfur in the interstellar medium.  The observation of a wide inventory of S-bearing species is also
necessary to better constrain the physical parameters, in particular the cosmic-ray ionization
rate for H$_2$ and its variation along the cloud.

\begin{acknowledgements}
 The research leading to these results has received funding from
 MICIU (Spain) under grants FIS2017-83473-C2, AYA2016-75066-C2-2-P, ESP2017-86582-C4-1-R, AYA2017-85111-P, PID2019-105552RB-C41  and PID2019-106235GB-I00.
 NB acknowledges the computing facilities by TUBITAK-TRUBA, and OR and AA 
 acknowledge computing time at Finisterre (CESGA) and Marenostrum (BSC)
 under RES computational grants ACCT-2019-3-0004 and  AECT-2020-1-0003.
 SPTM acknowledges the European Union’s Horizon 2020 research and innovation program
 for funding support under agreement No 639450 (PROMISE).
\end{acknowledgements}

%%%%%%%%%%%%%%%%%%%%%%%%%%%%%%%%%%%%%%%%%%%%%
%\bibliography{CSObib}

\end{document}